\documentclass[twocolumn, superscriptaddress, amsmath, amssymb, aps]{revtex4-2}

\usepackage{graphicx}
\usepackage{dcolumn}
\usepackage{bm}
\usepackage{xcolor}
\usepackage{mathtools}
\usepackage{physics}
\usepackage[normalem]{ulem}
\usepackage{soul}
\usepackage{braket}
\usepackage[T1]{fontenc}
\usepackage{xr}
\usepackage{makecell}

\usepackage{hyperref}
\hypersetup{colorlinks=true, citecolor=blue, linkcolor=blue, urlcolor=blue}

\begin{document}

\title{Unified Approach to Controlling Scattering in non-Hermitian Systems}

\author{Jared Erb}
 \email{jmerb@umd.edu}
 \affiliation{Maryland Quantum Materials Center, Department of Physics, University of Maryland, College Park, MD, 20742, USA}

\author{Nadav Shaibe}
 \affiliation{Maryland Quantum Materials Center, Department of Physics, University of Maryland, College Park, MD, 20742, USA}

\author{Steven M. Anlage}
 \affiliation{Maryland Quantum Materials Center, Department of Physics, University of Maryland, College Park, MD, 20742, USA}

\date{\today}

\begin{abstract}

The ability to controllably manipulate non-Hermitian wave scattering environments has been used to discover exotic scattering phenomena such as scattering exceptional points and coherent perfect absorption, and create numerous applications including signal routing, filtering, imaging, enhanced sensing, communication, wireless power transfer, etc. We introduce the concept that many of these applications and phenomena are fundamentally governed by singularities of the scattering matrix. With this understanding, our demonstrated ability to control the location of the scattering singularities themselves can be used to enhance current applications and develop new ones. In generic multiport complex scattering systems, there is an abundance of topologically protected scattering singularities corresponding to complex zeros of various functions of the scattering matrix. We show that with just three tunable parameters we are able to create the conditions for nearly any scattering singularity at arbitrary frequencies in such systems. With five tunable parameters, we demonstrate that more complex scenarios can be accomplished, such as making disparate singularities coincident in parameter space or placing singularities at different frequencies under the same system conditions. A benefit of systems with tunable parameters is that generic complex systems can be re-purposed to exhibit many different useful properties solely by reconfiguring their tunable parameters. A particularly interesting scattering phenomenon is coherent perfect absorption, where a specific wavefront injected into the system is completely absorbed, with no energy reflected or transmitted through any channel. After creating the conditions for this singularity at an arbitrary preselected frequency, we can find and \textit{inject} the coherent perfect absorption wavefront and demonstrate output-to-input power ratios as low as $1 \times 10^{-10}$, $2 \times 10^{-8}$, and $3 \times 10^{-8}$ in a quarter bow-tie microwave billiard with two, three, and four ports, respectively.

\end{abstract}


\maketitle

\section{Introduction}\label{sec.Intro}

Advancements in the understanding of wave phenomena and their implementation for applications are often enabled by new or improved means of controlling a scattering process. Modern approaches commonly use metasurface/cavity shaping and/or wavefront shaping techniques to exert control over a system  and achieve useful tasks. Metasurface/cavity shaping is the manipulation of the geometry and/or electromagnetic properties of a system to accomplish specific outcomes. Methods to control a system include the use of metamaterials and metasurfaces that are embedded in the scattering system \cite{Sievenpiper99,Cui14,Dupre15}. Wavefront shaping is the manipulation of the input signal to the system to achieve a specified goal, and in optics this can be accomplished through the use of a spatial light modulator \cite{Vellekoop07,Mosk12,Nixon13,Vellekoop15,Hougne16}. 

Non-Hermitian Hamiltonian systems have facilitated remarkable discoveries of exotic phenomena and their topological properties not found in their Hermitian counterparts \cite{Bender07,Gong18,Kawabata19,Ashida20}, such as the non-Hermitian skin effect \cite{Song19,Okuma20,Zhang22} and Hamiltonian exceptional points \cite{Berry2004,Lee08,Ryu12,Heiss12,Doppler16,Hodaei17,Alu19,Bergholtz21,Ding2022}. These discoveries have been extensively researched resulting in numerous insights and applications, and have laid the groundwork for similar studies to be done in other types of systems. Moving beyond non-Hermitian Hamiltonian systems, a growing body of research has been focused on studying complex wave scattering systems both theoretically \cite{Fyodorov97,Kottos00,Kottos03,Fulga11,Sounas17,Stone19,Sweeney20,Kang2021,Guo23,Ma25,Byrnes25,Alhulaymi25} and experimentally in many settings, including acoustics \cite{Song14,Shi16,Auregan16,Meng17,Richoux18}, microwaves \cite{Doron90,Kuhl2013,Pichler19,Zhao20,Chen20,Hougne21_2,Ferise22,Horodynski22,Wang24,Faul25,Erb25,Shaibe25}, optics \cite{Vellekoop07,Chong10,Lin11,Bruck13,Nixon13,Rotter17,Huang17,Baranov17,Guo17,Liu20,Sakotic21,Yang21,Kang22,Jiang2024,Guria24}, electric circuits \cite{Sakotic23}, etc. In this case one studies the scattering matrix ($S$), which incorporates the Hamiltonian, but has uniquely different non-Hermitian topological properties. For example energy/frequency becomes a parameter in the $S$-matrix, rather than the setting for eigenvalues and their degeneracies. The scattering matrix can be related to the Hamiltonian of the closed system through the Heidelberg approach \cite{VERBAARSCHOT1998,SOKOLOV1989,Fyodorov97,FSav11,Kuhl2013,Schomerus2015,Chen21}. With this approach, the non-Hermitian effective Hamiltonian is constructed from the Hamiltonian of the closed system and the matrix of coupling coefficients between the modes of the closed system and the scattering channels. Spatially uniform attenuation can be incorporated by replacing energy with a complex energy, and lumped attenuation can be accounted for by adding an additional matrix to the effective Hamiltonian which couples the modes of the closed system to localized absorbers which are treated as absorbing channels. Experimental scattering matrices are often non-unitary due to the presence of loss or gain in a system.

The scattering matrix relates incoming to outgoing waves of a system and is a ``blackbox'' approach which can be used to characterize systems where the internal wave interactions are unknown or too complex to simulate or model. Due to the ability to inject arbitrary wavefronts with tunable energy/frequency and phase, there exist new wave phenomena not seen in non-Hermitian Hamiltonian systems. However, the methods and tools developed for analyzing non-Hermitian Hamiltonian systems are also applicable to understanding scattering phenomena, such as topological properties and conservation laws.

The scattering matrix is often used to characterize linear wave scattering systems since it is directly measurable in many experimental contexts, from acoustics to optics. In this paper we are interested in generic complex scattering systems, which have complicated and irregular interior geometry with no symmetries and have overall linear dimensions that are much larger than the operational wavelength. The scattering systems have multiple tunable parameters $p$ that have an impact on the scattering properties, hence $S = S(p_1,p_2,...,p_n)$. These parameters include the frequency of the incident waves, as well as the displacement or rotation of perturbers, voltages applied to tunable metasurfaces that vary their reflection coefficient, variable attenuators or gain elements, geometrical modifications of the wave propagation region, etc. One can then examine the properties of the $S$-matrix in multi-dimensional parameter spaces. In particular, we will focus on scattering singularities, which arise when any complex scalar function of the scattering parameters achieves a value of $0+i0$, and therefore has an ambiguous phase at that point. Singularities of the scattering matrix are the foundation of many unique and exciting wave phenomena. These include reflectionless scattering modes \cite{Imani20,Sweeney20,Stone21,Sol2023,Jiang2024}, transmissionless scattering modes \cite{Kang2021,Huang2022,Faul25}, scattering exceptional points \cite{Erb25,Qin26,Xu26,Wang26}, coherent perfect absorption \cite{Chong10,Wan11,Baranov17,Chen20}, etc. Many applications are enabled by these phenomena, including imaging \cite{Hougne16,Slobodkin22}, sensing \cite{Zhang14,Hougne21_2,Sakotic21,Sakotic23}, filtering \cite{Faul25}, wireless power transfer \cite{Krasnok18,Hong26}, secure communications \cite{Imani20,Chen20}, energy routing \cite{Faul25,Alhulaymi25}, etc.

The parametric dependence of the scattering matrix has been extensively studied \cite{Beenakker97}, and it has been shown that scattering singularities are topologically stable \cite{Nye74,Liu20,Sakotic21,Sakotic23,Guo23,Dennis25,Erb25,Shaibe25_2,Alhulaymi25}. Due to their topological properties, the singularities can only be created or annihilated in pairs conserving the winding number for the phase of the complex scalar function that experiences a value of $0+i0$ \cite{Neu90,Sweeney20,Kang2021,Huang2022,Erb25}. A consequence of this conservation law is that small perturbations to the system only slightly modify the location of the scattering singularities, but do not destroy them. Additionally, even for generic complex scattering systems, in two-dimensional parameter spaces scattering singularities are relatively abundant \cite{Zhang10,Erb25,Shaibe25_2}. Therefore, in systems with tunable parameters, the location of scattering singularities can be controlled to a remarkable extent. For example, one can bring singularities to particular desired real frequencies, or disparate singularities can be forced into coincidence at a single real frequency. Due to the many practical applications enabled by scattering singularities, the ability to place them at arbitrary frequencies of interest can be extremely useful.

Recently, there has been much interest in reflectionless/transmissionless scattering modes (RSMs \cite{Imani20,Sweeney20,Stone21,Sol2023,Jiang2024}/TSMs \cite{Kang2021,Huang2022,Faul25}) and coherent perfect extinction (CPE) \cite{Guo23}, which are generalizations of coherent perfect absorption (CPA). CPA occurs when a specific wavefront injected into a scattering system though all ports is completely absorbed, with no reflection or transmission. RSMs, TSMs, and CPE describe scenarios where a particular input wavefront sent through a set of ports $I$ results in zero output on a set of ports $O$. RSMs occur when $I$ and $O$ are the same, TSMs occur when $I$ and $O$ are completely distinct, and CPE covers all cases. We define the size of $I$ as $j$, and the size of $O$ as $k$. When $j=k$, CPE is derived from a scattering singularity corresponding to the complex zero of the determinant of a particular subblock of the scattering matrix. In the simplest examples of CPE, the subblock is a single scattering matrix element. If $j>k$, CPE is always attainable, but if $j<k$, then CPE is more challenging to achieve but can be described as a coincidence of $(k-j+1)$ scattering singularities \cite{Guo23}. Alhulaymi et al. \cite{Alhulaymi25} extend the use of CPE occurring at complex frequencies and use the coincidence of CPEs to solve energy routing problems in multiport scattering systems. All of these approaches essentially utilize scattering singularities, whether they are just reflection or transmission zero singularities, the zero determinant of a subblock of the scattering matrix, or a coincidence of multiple scattering singularities. We shall adopt the language of scattering singularities on the real frequency axis to encompass all of these complementary approaches to control the properties of generic complex scattering systems.

The time delay of excitations sent through a scattering environment has also been extensively studied \cite{Wigner55,Smith60,Asano16,Chen21,Kang2021,Huang2022,Erb24}, where for non-Hermitian systems the time delay is complex. Complex time delays correspond to frequency/energy derivatives of complex scalar functions of the scattering parameters, and are related to time and frequency shifts of wave pulses traveling through a system \cite{Asano16,Giovannelli25,Giovannelli26}. Recent work has shown that every scattering singularity has a corresponding time delay that diverges as a simple pole at that point in parameter space \cite{Shaibe25,Shaibe25_2}. 

Many previous works have demonstrated impressive control over scattering singularities, but generally the scope of each work is limited in some way. Examples include consideration of only one specific scattering singularity, treating only one- or two-port systems, systems with special symmetry or geometry, lossless systems, or features studied through simulations only. The results in this work build upon these previous works, specifically Refs. \cite{Frazier20,Hougne21_2}, where they use programmable metasurfaces embedded in generic complex scattering systems to create scattering singularities at arbitrary frequencies, mainly focusing on coherent perfect absorption. Experimental demonstrations of controlling scattering singularities have often used metasurfaces with large numbers of individually addressable elements, but generally the change of each element of these metasurfaces is binary in nature and tends to have a very small effect on the scattering environment. Previous results have demonstrated the necessity of grouping multiple pixels of a metasurface together in order to meaningfully perturb the scattering system and to reduce the dimensionality of the tunable parameter space \cite{Frazier20,Hougne21_2,Frazier22}. Even with grouping multiple pixels together, having a discrete tunable parameter space can make it quite challenging to control scattering singularities compared to the use of continuously tunable parameters, such as those employed here.

Coherent perfect absorption has been demonstrated in many experimental contexts, but the majority of published work focus on only one- or two-port systems, often relying on imposed symmetries to achieve CPA \cite{Chong10,Pu12,Song14,Sun14,Wong16,Meng17,Baranov17,Chen20,Imani20,Frazier20,Erb24}. Recently more effort has been focused on demonstrating CPA in systems with three or more ports \cite{Li17,Zhang17,Richoux18,Pichler19,Hougne21,Hougne21_2,Guo23,Rontgen23,Faul25}, however most of these works only find the scattering singularity that enables CPA, and do not demonstrate the injection of the CPA wavefront into the system. References \cite{Slobodkin22,Horner24} have demonstrated the extraordinary ability to perform coherent perfect absorption for any arbitrary wavefront, however this requires a degenerate cavity which is highly specialized and not representative of the diverse scattering phenomena found in nature. Our approach is to investigate robust scattering singularities and extraordinary scattering phenomena that are found in generic complex scattering systems having little or no symmetry.

In this work, we experimentally demonstrate that multiport wave chaotic scattering systems with multiple tunable parameters can create nearly any scattering singularity at an arbitrary frequency within the tunability bandwidth of the parameters of the system. Any tunable parameter which can sufficiently affect wave scattering in the system can work. Our perspective is that numerous wave scattering phenomena and applications are enabled through appropriate manipulation of scattering singularities. Within this framework, we show that generic systems with tunable parameters can arbitrarily manipulate singularities, which enables a single system to exhibit many different scattering phenomena just by varying the tunable parameters. This allows a given system to perform many different practical applications or adapt to changes in a system environment over time. Lastly, we will also demonstrate the injection of a coherent perfect absorption wavefront at an arbitrary frequency in up to a four-port system, displaying over seven orders of magnitude of absorption, proving the feasibility of CPA for wireless power transfer or microwave heating applications in multiport systems.

\section{Experimental Setup} \label{sec.Setup}

The experimental system used in this work is a multiport quasi-two-dimensional quarter bow-tie microwave billiard, shown in the inset of Fig.\ \ref{Two_Sing_Manip} \cite{Stockmann90,Doron90,So95,Gokir98,Erb24,Erb26,Shaibe26}. The brass quarter bow-tie billiard has an area of 0.115 $m^2$ and a height of 7.9 mm, supporting quasi-two-dimensional electromagnetic modes when excited below 19 GHz. To manipulate the scattering environment there are multiple electronically controlled one-dimensional metasurfaces within the system. Each metasurface is 185 mm long, 7.9 mm high, and 1.8 mm thick, covering approximately 12\% of the perimeter of the billiard. The metasurfaces are comprised of 18 unit cells, where each subwavelength unit cell is a mushroom-shaped resonant element loaded with varactor diodes on two edges. Applying a global dc voltage bias to a metasurface tunes all diodes simultaneously, decreasing the capacitance of the diodes as the applied bias increases, thus changing the frequency-dependent properties of the metasurface and altering the amplitude and phase of reflected waves \cite{Sleasman23,Erb24,Erb25}. The operating frequency range of the metasurfaces is between 8 and 12 GHz. We label these tunable metasurfaces as $TM_q^{1D}$, where $q=1,2,3,...$ indicates each metasurface within the system. Additional details on the experimental system used can be found in Ref. \citenum{Erb26}. Although the results are demonstrated with this particular experimental setup, any generic multiport complex scattering system with multiple tunable perturbers has the potential to reproduce our results. It is not required for the system to have Lorentz reciprocity or any geometric or dynamical symmetries, but it must be operating under a linear regime, otherwise the scattering matrix is ill-defined.

For all experimental results, we start by measuring an $M \times M$ scattering matrix $(S)$ using a Keysight N5242B microwave vector network analyzer, where $M$ is the number of single-mode ports or channels connected to the scattering system. The scattering matrix relates the incoming and outgoing wave excitations for each port. We measure $S$ as a function of frequency and metasurface applied bias voltages, where the voltage biases are supplied using Keithley 2230G-30-1 and Rigol DP932A triple-channel programmable dc power supplies. The $S$-matrix measurements are carried out at low microwave power ($<0$ dBm) where the response of the metasurface-loaded cavity is linear.

\section{Results} \label{sec.Results}
\subsection{Scattering Singularities On Demand}

There are infinitely many scattering singularities that can be defined as the zero of a complex scalar function of the scattering parameters, such as products or polynomials of the scattering matrix elements, but the simplest examples are when individual elements of the scattering matrix go to zero (for example $S_{11}=0+i0$ or $S_{21}=0+i0$). In generic complex systems, scattering singularities have been shown to be topologically protected and relatively abundant in two-dimensional parameter spaces \cite{Liu20,Sakotic21,Sakotic23,Guo23,Erb25,Shaibe25_2,Alhulaymi25}. Therefore, using the tunable metasurfaces within our system, it is possible to move singularities to the real frequency axis at any frequency of interest $f_i$. Applying an optimization algorithm to this task can make the positioning of any singularity fast and efficient. We use a surrogate optimizer prebuilt in MATLAB's global optimization toolbox to control the values of the applied bias voltages to the metasurfaces, but any other applicable optimizer should work. The absolute value of a complex scalar function corresponding to a particular scattering singularity ($|S_{21}(f_i)|$ for example) is the objective function to minimize each iteration. As we will never truly reach a value of zero for a scattering singularity, we end the optimization when the objective function is below an acceptable threshold. In a passive lossy system it is impossible for any scattering matrix element to have a magnitude greater than unity, but unconventional singularities that are the zeros of complex scalar functions such as $S_{11}S_{33} - 0.5S_{22}$, $Tr(S)-2det(S)$, or $S_{21} - (0.5-0.24i)$ are equally valid scattering singularities. However, the possible existence of scattering singularities does not guarantee that the optimization process will be able to find the necessary conditions to exhibit them.

\begin{table}[htb]
    \centering
    \renewcommand{\arraystretch}{1.5}
    \setlength{\tabcolsep}{0.5pt}
    \begin{tabular}{|c|c|c|c|c|}
    \hline
    \multicolumn{5}{|c|}{\textbf{Preselected Frequency = 9.197 GHz}}  \\ \hline
    \textbf{Singularity} & \textbf{Value (dB)} & \textbf{$\bm{TM_1^{1D}}$(V)} & \textbf{$\bm{TM_2^{1D}}$(V)} & \textbf{$\bm{TM_3^{1D}}$(V)} \\    \hline
    $S_{11}$ & -100.83 & 0.0543 & 0.5624 & 5.4380 \\    \hline
    $S_{22}$  & -85.68 & 11.3581 & 11.5289 & 4.2245 \\    \hline
    $S_{33}$ & -104.21 & 6.3667 & 2.7061 & 1.3081 \\   \hline
    $S_{21}$ & -85.19 & 4.1772 & 11.5964 & 7.0946 \\   \hline
    $S_{31}$ & -84.82 & 7.7458 & 7.0279 & 3.3095 \\   \hline
    $S_{41}$ & -87.35 & 1.1856 & 0 & 8.1741 \\   \hline
    $S_{32}$ & -89.82 & 9.8139 & 5.5453 & 0.4920 \\   \hline
    $S_{42}$ & -84.99 & 4.9468 & 0.9104 & 2.9559 \\   \hline
    $S_{43}$ & -86.71 & 10.7415 & 2.0550 & 0 \\   \hline
    $det(S)$ & -105.97 & 6.9558 & 1.8908 & 3.9663 \\   \hline
    $S_{22}-S_{33}$ & -86.60 & 10.2675 & 0.0342 & 4.3939 \\   \hline
    $S_{22}-S_{32}$ & -85.24 & 0.3750 & 2.0494 & 4.7940 \\   \hline
    $S_{42}-S_{31}$ & -93.35 & 1.1689 & 12 & 9.3207 \\   \hline
    $S_{44}-S_{12}$ & -86.97 & 0.7663 & 0 & 0.4123 \\   \hline
    $S_{23}-S_{14}$ & -87.66 & 4.8265 & 0 & 8.9919 \\   \hline
    $S_{22}-S_{44}$ & -89.34 & 4.2215 & 1.5794 & 11.7182 \\    \hline

    \end{tabular}
    \caption{Table of scattering singularities created at an arbitrarily predetermined frequency of 9.197 GHz in the four-port quarter bow-tie microwave billiard. The first column indicates the specific singularity, the second column shows the measured value of the singularity (in dB) after the optimization process is complete, and the last three columns show the voltage values applied to the three metasurfaces required to exhibit the scattering singularity.}
    \label{Sing_Opt_Table}
\end{table}

In Table \ref{Sing_Opt_Table}, using three metasurfaces along with the optimization process, we show that we can find conditions to exhibit multiple unique singularities at an arbitrarily selected frequency of 9.197 GHz in the four-port quarter bow-tie billiard. We demonstrate the zero of all scattering matrix elements except for $S_{44}=0+i0$ due to poor coupling of the fourth antenna at this frequency. Only half of the off-diagonal $S$-matrix elements are explicitly shown as the quarter bow-tie billiard is a reciprocal system, meaning that $S_{xy} = S_{yx}$. We stopped at the creation of 16 scattering singularities as we believe that amount sufficiently demonstrates that scattering singularities are easily created in a generic system with multiple tunable parameters.

For most of the examples in Table \ref{Sing_Opt_Table}, all three metasurfaces were tuned to find the singularity conditions, although for some examples such as $S_{41}$, only two of the metasurfaces needed to be varied. Nominally, two independent tunable parameters should be sufficient to find the conditions for any singularity since complex scalar functions are codimension 2 \cite{Guo23,Alhulaymi25}, but the more tunable parameters available, generally the faster and easier it is to achieve any given condition. In theory, to arbitrarily manipulate $N$ singularities, only $2N$ tunable parameters are required. In practice, tunable parameters usually don’t have fully independent effects on the scattering singularities, so more than $2N$ tunable parameters are needed. The more singularities that are being manipulated, the longer the optimization process generally takes and the more complicated the optimization landscape becomes, which makes finding global minima potentially challenging.

With the ability to manipulate multiple singularities, each singularity can be placed at a different frequency or independent singularities can be combined at the same frequency. We call the combination of multiple singularities at the same frequency `composite singularities'. A useful case of a composite singularity can be demonstrated with energy routing, where for a specific frequency, energy entering one port of a system only leaves in a separate port, such that there is no reflection or transmission to other ports \cite{Faul25,Alhulaymi25}. In a four-port system where the energy from port 1 is intended to only exit port 3, this example would be the combination of an $S_{11}=0+i0$ reflection zero singularity and $S_{21},S_{41}=0+i0$ transmission zero singularities all at the frequency of interest. The optimization process to control the location of $N>1$ singularities is almost exactly the same as it was for one singularity. The only thing that changes is the objective function, where we define a sum of absolute values of each complex scalar function at the frequency of interest for each singularity (for example $|F(S(f_1))| + |G(S(f_2))| + ... + |H(S(f_N))|$). After the conditions for the singularities have been established, small perturbations to the system can affect each singularity differently, making the reestablishment of the original conditions more complex than for the single singularity case.

\begin{figure}[htb]
\centering
\includegraphics[width=8.9cm]{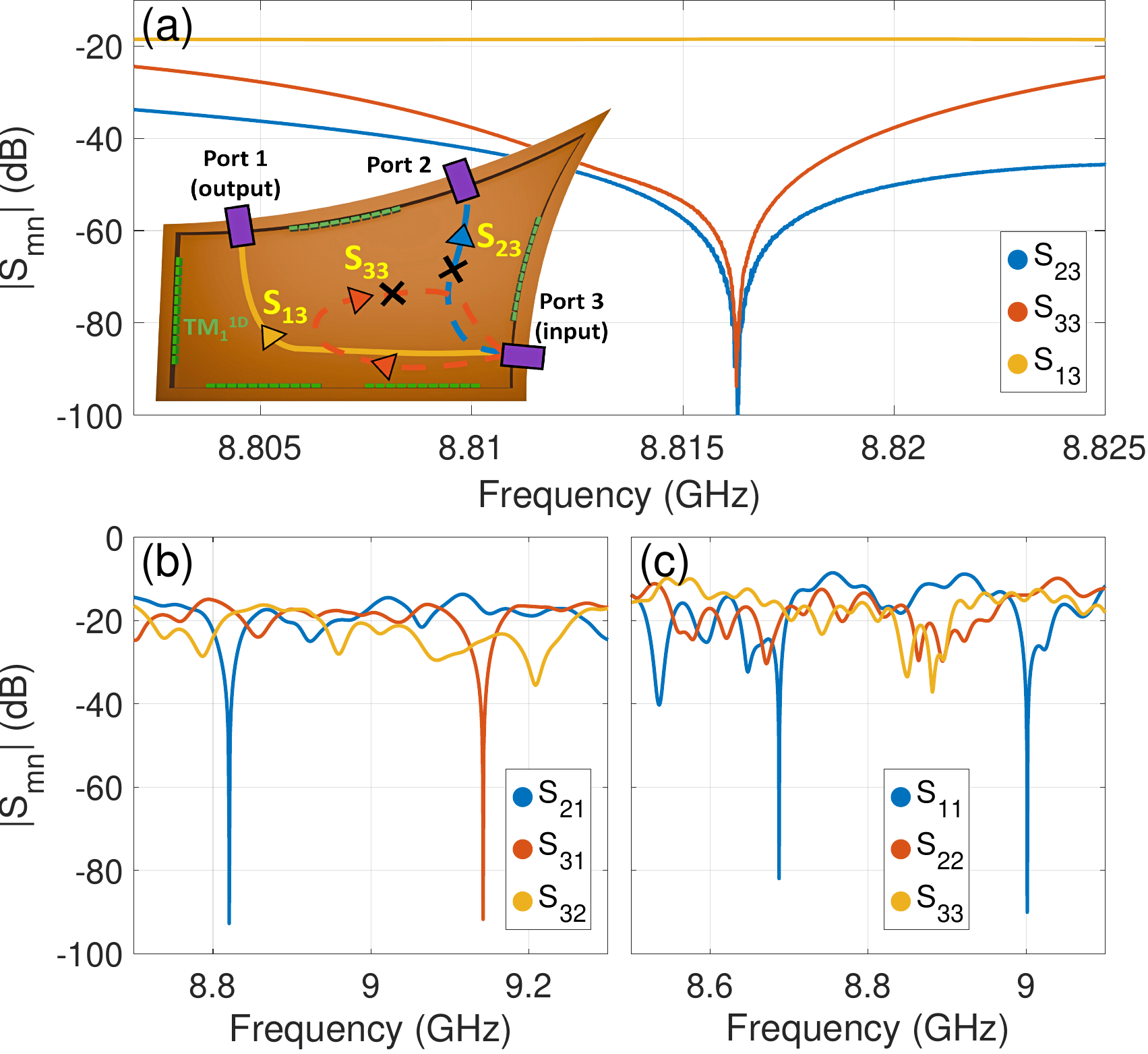}
\caption{(a) Three scattering matrix elements vs frequency after an optimization process to create a composite singularity of two of the matrix elements, demonstrating energy routing from port 3 to port 1 at the frequency of the minimum. The inset shows a schematic of energy routing from port 3 to port 1 in the quarter bow-tie billiard. (b,c) Three scattering matrix elements vs frequency after an optimization process to create two singularities existing at the same cavity conditions, but at different frequencies. All of these measurements were taken in the quarter bow-tie billiard with five metasurfaces (green strips in the inset) used in the optimization process. The leftmost metasurface is labeled $TM^{1D}_{1}$, and the rest of the metasurfaces ($TM^{1D}_{2-5}$) are clockwise ordered.}
\label{Two_Sing_Manip}
\end{figure}

In Figure \ref{Two_Sing_Manip}, we demonstrate that using five metasurfaces we can arbitrarily manipulate two singularities in the three-port quarter bow-tie billiard. In Figure \ref{Two_Sing_Manip}(a), we demonstrate the conditions for a composite singularity, which is a routing example from port 3 to port 1 where there is nearly 70 dB of discrimination. In Figure \ref{Two_Sing_Manip}(b,c) we find the conditions to simultaneously create two singularities at different frequencies. For the examples shown in Figure \ref{Two_Sing_Manip}, we allowed the frequency to be tuned in a small range around the chosen frequencies of interest to reduce the optimization time. In practice, this allows frequency to also be a tunable parameter of the system.

\subsection{Multiport Coherent Perfect Absorption}

For coherent perfect absorption to be possible, the scattering system must satisfy the $det(S)=0+i0$ condition (equivalently a zero $S$-matrix eigenvalue) at a particular frequency, which requires the system to have some amount of loss. The determinant of the scattering matrix is a complex scalar function of codimension 2 for systems with any number of ports. Therefore achieving a $det(S)=0+i0$ singularity is generically possible for any complex scattering system as long as there are at least two tunable parameters, regardless of the number of ports. However, in systems with a small amount of loss and a large number of ports, or systems with poor antenna coupling, achieving CPA isn't guaranteed to be possible. In Appendix \ref{sec.A}, the simultaneous placement of two $det(S)=0+i0$ singularities at different frequencies was demonstrated using the optimization process. For potential practical applications, multiple $det(S)=0+i0$ conditions can be placed nearby in frequency, enhancing the level of absorption around the conditions, or the conditions could be placed at particular frequencies of interest, allowing CPA to be demonstrated at multiple frequencies under the same system conditions if the input signals have multiple frequency components.

\begin{figure}[htb]
\hspace*{-0.2cm}
\centering
\includegraphics[width=7.0cm]{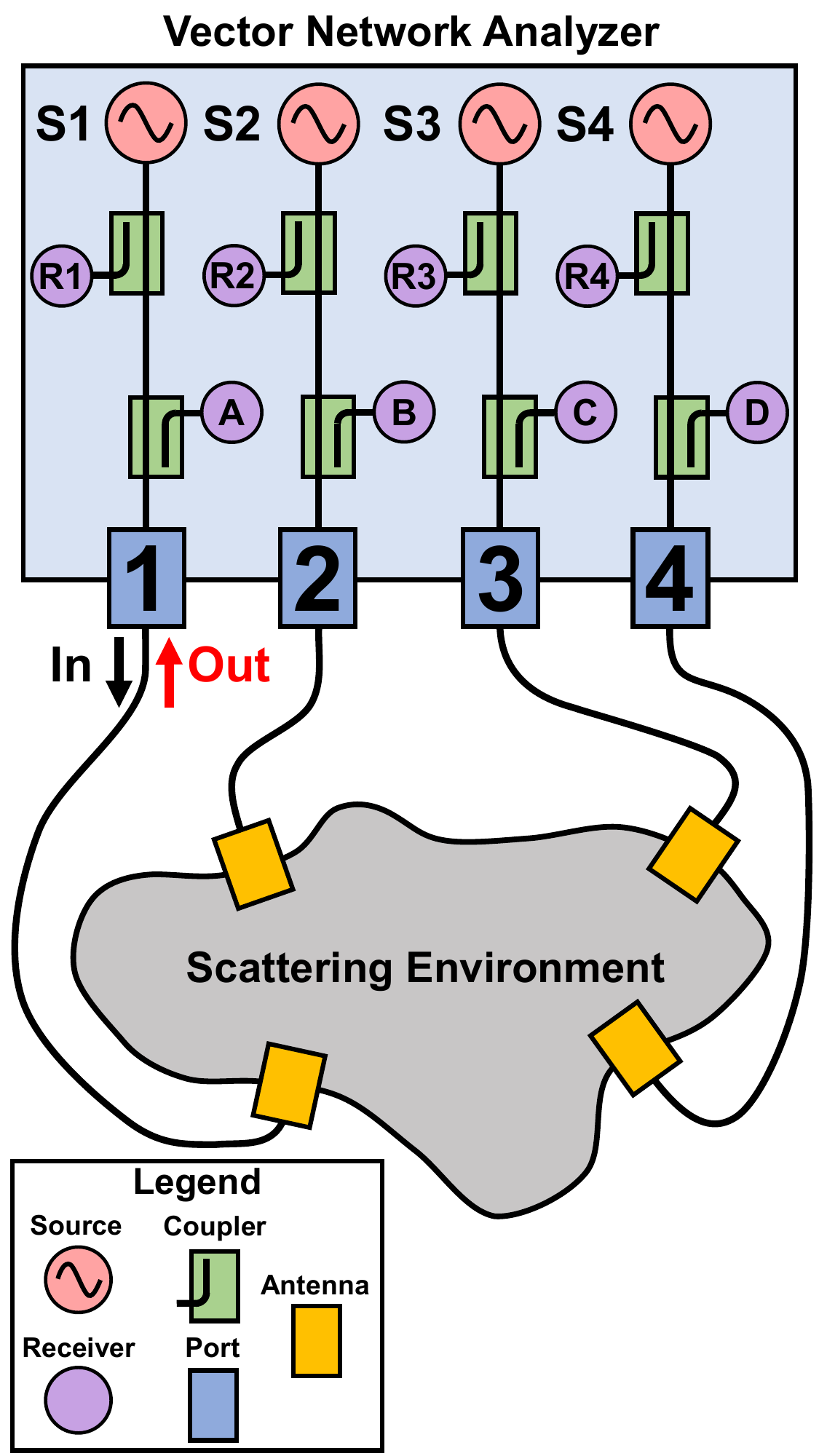}
\caption{Schematic of the vector network analyzer in a multi-source injection setup connected to a scattering system. The input wavefront injected into the four-port system from the independent sources (S1-S4) is measured with the receivers R1-R4 through couplers, and the output signal returning from the system is measured with the receivers A-D through couplers. The receivers measure both the magnitude and phase of the signals.}
\label{Receiver_Schematic}
\end{figure}

To exhibit coherent perfect absorption, the eigenvector corresponding to the zero eigenvalue must be injected into the system. For an $M$-port scattering system, this requires independent control of $M$ sources that can be routed through the ports used to find the $det(S)=0+i0$ condition. The measurement setup of injecting an arbitrary signal into the system is quite a bit different from the previous results which were from measurements of the scattering matrix. In an injection setup, receivers inside the vector network analyzer (VNA) measure the amplitude and phase of both the ingoing and outgoing signals to and from the system. A schematic illustrating the injection measurement for a four-port system is shown in Figure \ref{Receiver_Schematic}. In the schematic, the VNA was based on the four-port test set block diagram in the technical specification document for the Keysight N5242B PNA-X Microwave Network Analyzer. The amplitude and phase of the independent microwave sources (pink circles labeled S1-S4) can be controlled to create an arbitrary input wavefront. The receivers measuring the ingoing signals (purple circles labeled R1-R4) through couplers (green boxes) can be used to verify that the desired wavefront is created. The wavefront is sent to the scattering system through the ports (blue boxes labeled 1-4) and coaxial cables (black curves) connecting to the antennas (yellow boxes). After the wavefront interacts with the scattering system and returns back to the VNA, the receivers A-D measure the outgoing signals.

For a calibrated scattering matrix measurement, the measurement reference plane is shifted to the plane of calibration (usually to the end of the coaxial cables that connect to the scattering system), removing cable effects and any internal path differences inside the network analyzer. For an injection setup, calibrations cannot be applied. The measurement reference plane is required to be at the signal receivers inside the network analyzer, which directly measure the ingoing signals to the system and the outgoing signals from the system. Additionally, the independent sources need to be routed through the ports, changing the internal paths the signals travel. Therefore, when taking scattering matrix measurements to find $det(S)=0+i0$ conditions for CPA injection, we leave the system uncalibrated, keeping the measurement reference plane at the receivers. After the CPA conditions are found and the system is switched to an injection setup, the independent sources can be adjusted using information from the receivers until the relative inputs (magnitude and phase) match the CPA eigenvector. To quantify the degree to which coherent perfect absorption is achieved, we use the output to input power ratio measured by the receivers defined as $\sum_{i=1}^{M} P_{out,i} / \sum_{i=1}^{M} P_{in,i}$, where $P_{out,i}$ ($P_{in,i}$) is the output (input) power measured at port $i$.

In practice, adjusting the independent sources so that the receivers measure a particular wavefront is non-trivial. Changes in the source's power and phase generally aren't identical to the change the receiver measures, and each source's power and phase typically are not completely independent, so adjusting one will alter the other. To achieve a high degree of absorption, the relative amplitudes and phases between each ingoing receiver need to precisely match the CPA eigenvector. The simplest method to match the CPA eigenvector is to use a similar optimization process used to find $det(S)=0+i0$ conditions, but in this case the optimizer controls the amplitude and phase of each independent source, and the objective function to minimize is the overall output power from the scattering system, as  measured from the outgoing receivers. To obtain a good initial guess for the optimization, the power and phase of one of the sources can be fixed, and then the other sources can be straightforwardly adjusted relative to the fixed source based on the relative components of the CPA eigenvector.

\begin{figure}[htb]
\centering
\includegraphics[width=8.7cm]{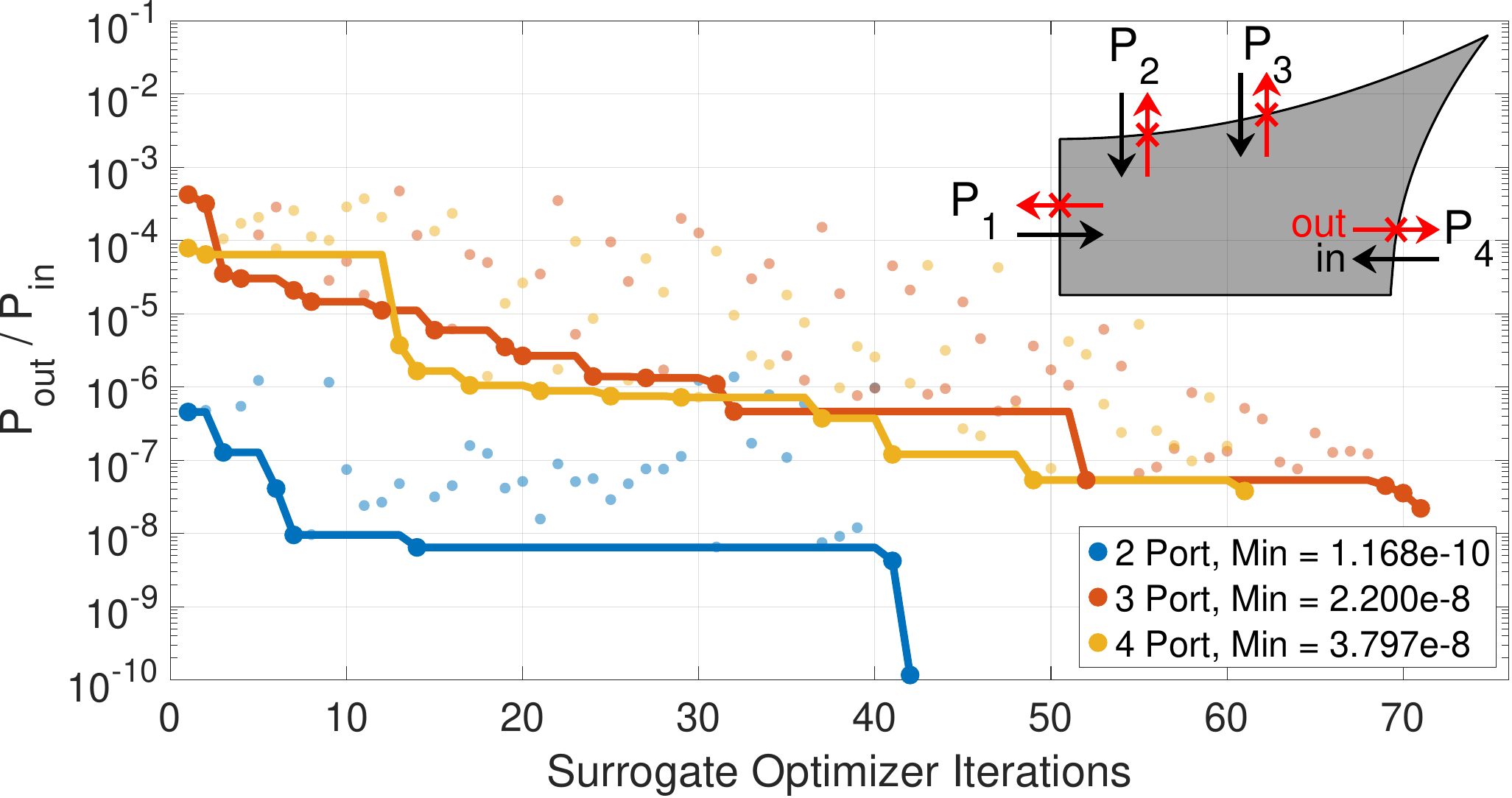}
\caption{Coherent perfect absorption (CPA) wavefront injection optimization process in the quarter bow-tie billiard. The blue points correspond to the value of the output to input power ratio at each optimization step for a two-port injection. The orange and yellow points have the same meaning as the blue points, but for a three- and four-port injection respectively. The solid lines follow the cumulative minimum value through the optimization iterations. For each of these measurements we first created $det(S)=0+i0$ conditions at an arbitrary frequency using the singularity ``on demand'' optimization process. The frequencies of the CPA points for the two through four-port systems were 8.235 GHz, 9.146 GHz, and 9.378 GHz, respectively. The inset is a schematic of a four-port CPA injection, where the ingoing (outgoing) signals are marked with black (red) arrows.}
\label{Multi_Port_CPA}
\end{figure}

In Figure \ref{Multi_Port_CPA}, we demonstrate coherent perfect absorption in the quarter bow-tie billiard with two, three, and four ports. The output to input power ratio $P_{out}/P_{in}$ is plotted versus each iteration of the injection optimization process, and the solid lines follow the cumulative minima throughout the iterations. See Appendix \ref{sec.A} for a table containing the individual input and output power values measured by the receivers at the beginning and end of each of the CPA wavefront injection optimizations shown in Figure \ref{Multi_Port_CPA}. For systems with two or more ports, coherent perfect absorption is a doubly singular phenomenon, requiring both a $det(S)=0+i0$ condition and a perfect matching of the injected CPA eigenvector, so we will never truly reach complete absorption experimentally. But for each case, we have over seven orders of magnitude of absorption of the input signals at the minimum. In each demonstration, we first found the $det(S)=0+i0$ condition at an arbitrary frequency using the optimization process described to make a singularity ``on demand''. Then we switched the network analyzer to an injection setup and used the injection optimization process to minimize the output power using the eigenvector corresponding to the smallest eigenvalue as a good initial guess of the CPA eigenvector.

Coherent perfect absorption can be demonstrated in any lossy generic complex scattering system with tunable parameters. A single-port CPA demonstration is fairly simple since this only requires the system to be at a reflection zero singularity (RSM), then any signal sent into the system is almost entirely absorbed. In Figure \ref{One_Port_CPA}, we demonstrate this for input signals of varying power and phase in the one-port quarter bow-tie billiard. In Figures \ref{1D_3_Port_CPA} and \ref{3D_2_Port_CPA}, coherent perfect absorption is demonstrated using a three-port one-dimensional microwave graph with tunable phase shifters and a two-port three-dimensional microwave cavity with two-dimensional metasurfaces, respectively. Along with these figures, we have shown one-port through four-port CPA accomplished at arbitrary frequencies in three generic experimental systems with different wave propagation dimensions and unique tunable parameters.

\section{Discussion}\label{sec.Disc}

The main factor ultimately limiting the degree to which we can demonstrate scattering singularities is the noise floor of the network analyzer, see Appendix \ref{sec.B} for a characterization of the noise floor. Although in practice we generally stopped the optimization when the objective function dropped below our acceptable threshold of -80 dB, which is sufficiently small yet still above the noise floor of the VNA. We were also limited to manipulating up to two scattering singularities simultaneously as our system only has five tunable parameters. Other limiting factors include the resolution of the voltage step size of the power supplies used to bias the metasurfaces, the stability of the system environment over time, including the temperature, humidity, and atmospheric pressure of the room, and the time allowed for the optimization process to operate, which can depend on the number of tunable parameters and complexity of the objective function. For other experimental systems, the impact of the tunable parameters can be a significantly limiting factor. The tunable parameters need to impart a meaningful change on the scattering environment.

The VNA used in this work has four ports and two internal sources, limiting us to demonstrating CPA in at most a four-port system. The two additional independent microwave sources we used are Keysight model N5183B microwave analog signal generators, which are fed into two ports of the VNA to allow four independently controlled sources. Achieving CPA in higher port-number systems is relatively straightforward following the procedures described earlier, but will require equipment able to measure a high port-number system and be able to route independent microwave sources through each of the ports. The more ports that are used, the more challenging and time-consuming matching all the relative components of the CPA eigenvector can become.

There are multiple factors that limit our ability to achieve true complete absorption. These include how close to complex zero we achieved the $det(S)=0+i0$ conditions, which depend on the same limitations for demonstrating any scattering singularity, the stability of the microwave sources over time, and the noise floor of the network analyzer receivers which ultimately limit how small $P_{out}$ can get. The receiver noise floor and stability of the microwave sources and the limits that they impose on our ability to achieve CPA are discussed in Appendix \ref{sec.B}. The maximal value of $P_{in}$ is also limited to a value that maintains the varactor diodes of the metasurfaces in the linear regime. If the input power is too high, the varactor diodes become nonlinear, changing the scattering environment compared to when the $det(S)=0+i0$ conditions were found. In our experimental setup, the metasurfaces generally started behaving nonlinearly around +5 dBm of input power. The three- and four-port CPA demonstrations have a similar minimum value approximately two orders of magnitude larger than the two-port CPA. This is mainly caused by the two external microwave sources, which are unable to share a common local oscillator with the internal sources, causing significant phase drifting which limits our ability to reduce the output-to-input power ratio, as shown in Fig \ref{Phase_Noise}.

At the end of the CPA wavefront optimizations in Figure \ref{Multi_Port_CPA}, the output powers measured by the receivers were not in the noise floor, see Table \ref{CPA_Opt_Table}. For our setup, each iteration of the two-port wavefront optimization takes approximately 0.8 seconds, while the three- and four-port iterations take approximately 1.5 seconds. During the finite duration of the optimization process, the input amplitudes/phases of the microwave sources can independently vary, and the system itself can slightly vary. These variations reduce the quality of the optimization process which assumes a fixed system and stable parameters. We believe these variations are the dominant limiting factor for our demonstration of coherent perfect absorption. At extremely small values of the output to input power ratio, a small change to the system (related to the $det(S)=0+i0$ condition) or input wavefront relative values can significantly increase the amount of output power leaving the system. This can be seen in plots where a single parameter is varied through its optimal value corresponding to the CPA condition, while all other parameters are fixed at their optimal CPA value, see Figure 6 of Ref. \cite{Erb24}.

Coherent perfect absorption has the potential to greatly enhance applications in wireless power transfer, sensing, and microwave heating. For wireless power transfer and microwave heating, implementing CPA has the potential to significantly increase the efficiency of energy deposition to the target compared to conventional approaches \cite{Lerosey06,Fager14,Gowda16,Cangialosi16,Krasnok18,Neumaier19,Yang23}. Additionally, due to our control over the CPA conditions along with the topological protection of the $det(S)=0+i0$ scattering singularity, if heating (for example) causes a target to change its properties over time, and therefore the scattering environment, an optimization algorithm such as a feedback loop could adjust the cavity conditions and/or input signals to continuously maintain the CPA condition. This feedback method can also be applied to any other scattering singularities for any small perturbations to the system.

All of the demonstrated results are applicable to any linear wave scattering system, regardless of the presence of Lorentz reciprocity or geometric/dynamical symmetries. In this work the scattering systems remained linear, but for nonlinear systems a different approach would likely be needed to controllably manipulate scattering singularities. In particular, coherent perfect absorption has been demonstrated in nonlinear systems, enhancing the applicability of CPA, such as increasing the absorption over a broad range of frequencies \cite{Li18,Mullers18,Zezyulin18,Wang21,Suwunnarat22,Cui24,Wang24}.

\section{Conclusion}\label{sec.Conc}

Numerous wave scattering phenomena and applications are fundamentally governed by singularities of the scattering matrix. By demonstrating the ability to arbitrarily control scattering singularities in generic complex systems containing tunable parameters, we show that a given system can exhibit many different scattering phenomena. These phenomena include sensing, filtering, communication, energy routing, wireless power transfer, imaging, etc. We believe that viewing scattering problems through the approach of manipulating scattering singularities can be used to enhance current practical applications and develop new ones.

Beyond solely manipulating scattering singularities, we have also demonstrated wavefront shaping applications at arbitrary frequencies, showing that complex wave injection phenomena can also be accomplished in generic systems. In demonstrating coherent perfect absorption, we combined both cavity- and wavefront-shaping to first create $det(S)=0+i0$ conditions at a particular frequency, then to adjust the input signal until the CPA wavefront was matched. Using optimization processes to accomplish these goals, we demonstrated coherent perfect absorption in multiport systems with an absorption greater than seven orders of magnitude compared to the input signal.

\section*{Acknowledgments}
This work was supported by NSF/RINGS under grant No. ECCS-2148318, ONR under grant N000142312507, and DARPA WARDEN under grant HR00112120021.


\appendix

\section{Coherent Perfect Absorption in Multiport Generic Complex Scattering Systems}\label{sec.A}

\begin{figure*}[htb]
\hspace*{-0.05cm}
\centering
\includegraphics[width=18.0cm]{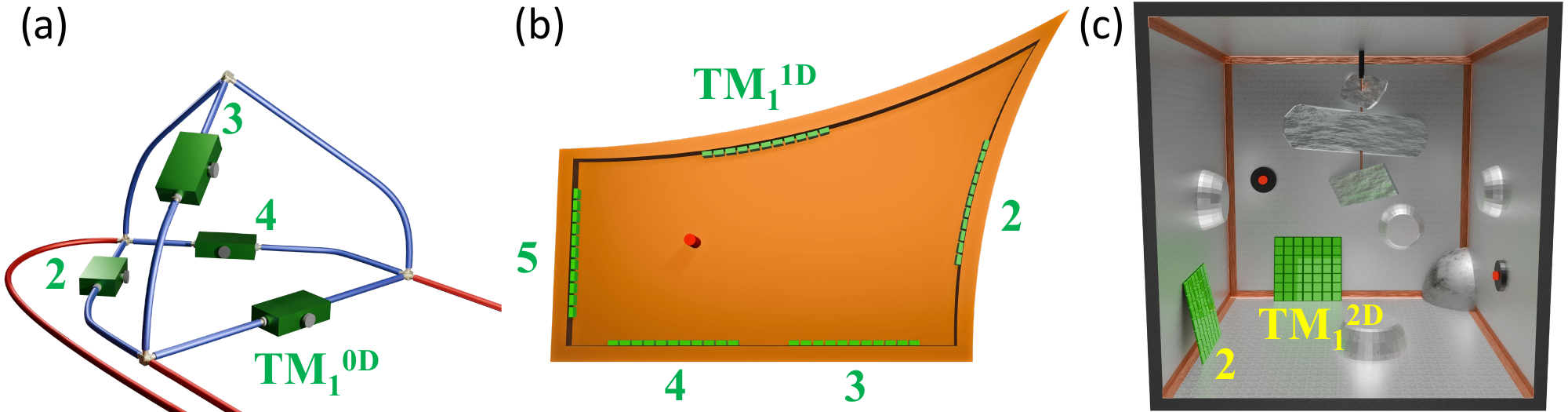}
\caption{(a) Schematic view of a three-port quasi-one-dimensional microwave graph. (b) Schematic view of a one-port quasi-two-dimensional quarter bow-tie billiard with the top lid removed. (c) Schematic view of a two-port three-dimensional microwave cavity along with a mode-stirrer and irregular scatterers, with the front wall removed. The tunable parameters within each system are the green objects labeled by $TM_q^{rD}$, where $q$ labels each device within a system and $r$ indicates the dimension of the device. For visual clarity only $TM_1^{rD}$ is fully labeled, while the rest of the tunable parameters within the systems are indicated by their $q$ value. The ports connecting the vector network analyzer to each system are indicated by the red objects.}
\label{All_Exps}
\end{figure*}

In this appendix, a figure showing the simultaneous creation of two $det(S)$ zero singularities at two different frequencies is shown. A table containing the measured input and output powers for the beginning and end of the coherent perfect absorption (CPA) wavefront injection optimizations in Figure \ref{Multi_Port_CPA} is shown. Coherent perfect absorption is also demonstrated in three generic complex scattering systems of different wave propagation dimensions, see Figure \ref{All_Exps}. The demonstrations are of one-port CPA in the quarter bow-tie billiard, three-port CPA in a microwave graph, and two-port CPA in a three dimensional cavity, respectively. Demonstrations of the CPA injection optimization process for randomly initialized input wavefronts and the average output to input power ratio over many random wavefronts are also shown.

\begin{table*}[htb]
    \centering
    \renewcommand{\arraystretch}{1.8}
    \setlength{\tabcolsep}{5pt}
    \begin{tabular}{|c|c|c|c|c|c|c|c|c|}
    \hline
    & \multicolumn{8}{c|}{\textbf{Receiver Power Values (dBm)}} \\ \cline{2-9}
    & \textbf{R1} & \textbf{R2} & \textbf{R3} & \textbf{R4} & \textbf{A} & \textbf{B} & \textbf{C} & \textbf{D} \\    \Xhline{2.5pt} \hline
     Two Port CPA Optimization Start & -5.133 & N/A & -19.706 & N/A & -83.455 & N/A & -68.533 & N/A  \\    \hline
     Two Port CPA Optimization End & -5.119 & N/A & -19.637 & N/A & -107.422 & N/A & -107.186 & N/A  \\     \Xhline{2.5pt} 
     Three Port CPA Optimization Start & -8.070 & -18.300 & -2.639 & N/A & -37.167 & -41.145 & -44.649 & N/A  \\    \hline
     Three Port CPA Optimization End & -7.660 & -18.996 & -2.636 & N/A & -84.723 & -81.331 & -82.749 & N/A  \\    \Xhline{2.5pt} 
     Four Port CPA Optimization Start & -2.517 & -9.219 & -4.904 & -25.851 & -44.346 & -47.095 & -46.712 & -57.918  \\    \hline
     Four Port CPA Optimization End & -2.510 & -8.947 & -4.555 & -25.601 & -82.365 & -80.381 & -82.975 & -77.090  \\    \hline
    
    \end{tabular}
    \caption{Table of power values measured by the input and output receivers at the start and end of the coherent perfect absorption wavefront optimizations shown in Figure \ref{Multi_Port_CPA}. The first column indicates the number of ports used in the optimization and whether it is for the start or end of the optimization. The rest of the columns correspond to the power in dBm measured by the input and output receivers, following the labeling convention of Figure \ref{Receiver_Schematic}. A value of N/A means that the receiver was not used for that optimization.}
    \label{CPA_Opt_Table}
\end{table*}

\begin{figure}[htb]
\hspace*{-0.28cm}
\centering
\includegraphics[width=8.9cm]{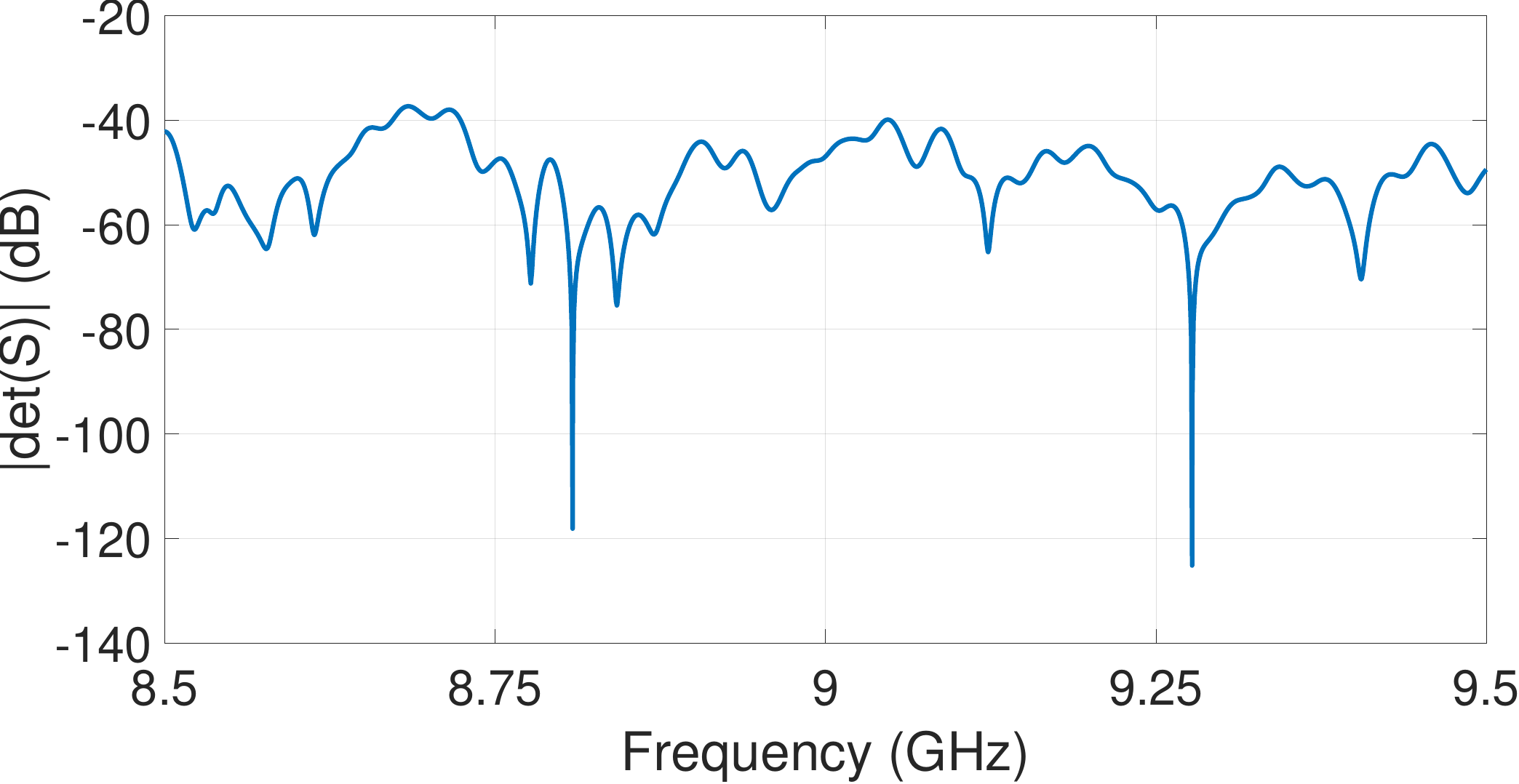}
\caption{$|det(S)|$ vs frequency after an optimization process to create two $det(S)$ zero singularities at different frequencies under the same cavity conditions. This was measured in the four-port quarter bow-tie billiard with five metasurfaces used in the optimization process (see inset of Fig. \ref{Two_Sing_Manip}(a)).}
\label{Two_DetS_Sing}
\end{figure}

In Figure \ref{Two_DetS_Sing}, we show that for the four-port quarter bow-tie billiard we are able to create $det(S)=0+i0$ conditions at two separate frequencies simultaneously using five tunable metasurfaces along with the optimization process. The ability to control the locations of any two singularities can be done in generic scattering systems with at least four independent tunable parameters, with a few more examples shown in Figure \ref{Two_Sing_Manip}.

In Table \ref{CPA_Opt_Table}, we show the input and output power values measured by the receivers for the beginning and end of each of the CPA wavefront optimizations shown in Figure \ref{Multi_Port_CPA}. Over the course of the optimization, the change in input powers (measured by receivers R1-R4) is quite small, indicating that using the eigenvector corresponding to the near zero eigenvalue as the initial guess is a good strategy. In contrast, the output powers (measured by receivers A-D) can drop by as much as 47 dBm from the beginning to the end. The overall output power is generally dominated by the largest component, therefore over the course of the optimization the initial highest power output generally drops the most.

\begin{figure}[htb]
\centering
\includegraphics[width=8.9cm]{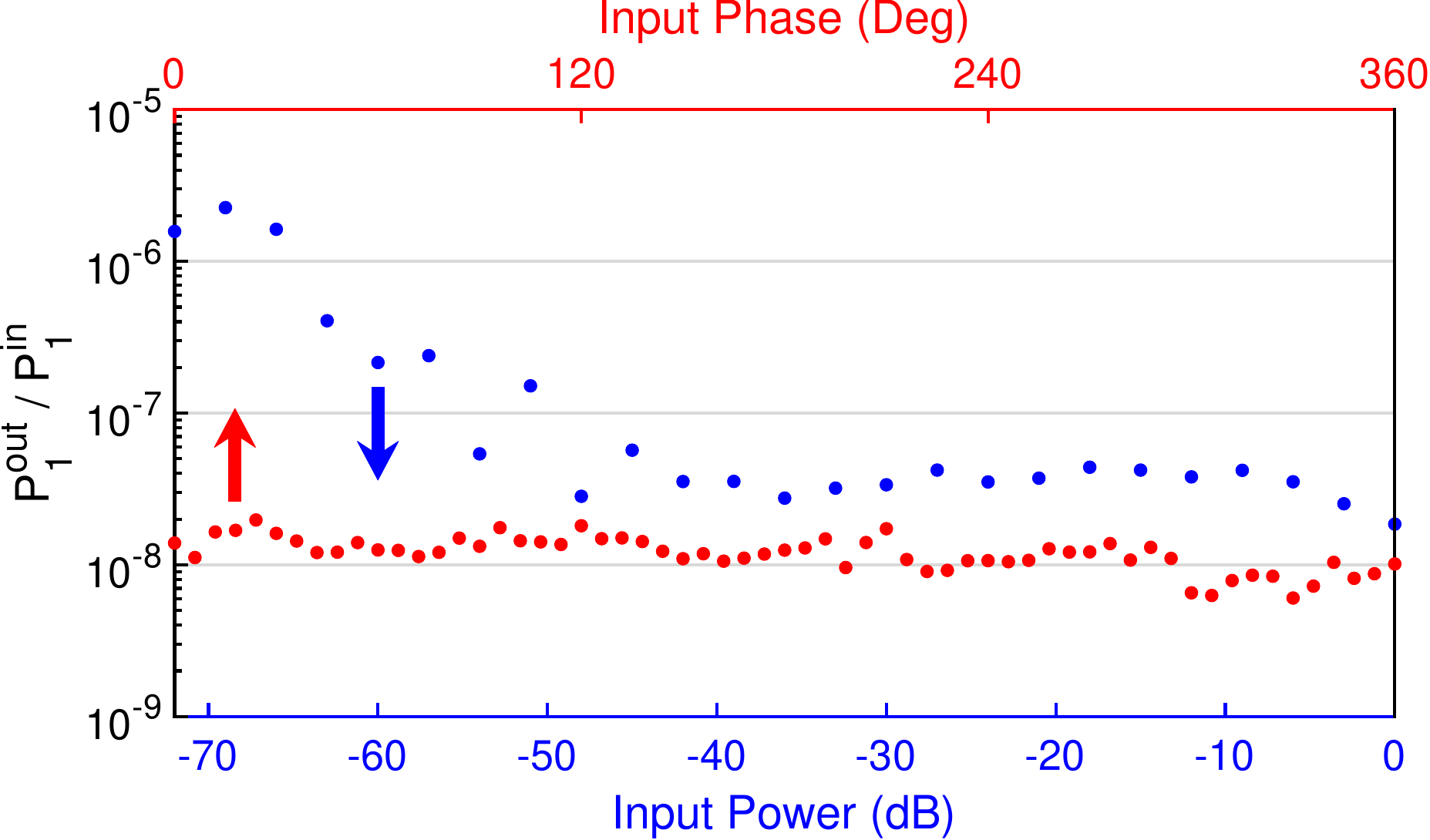}
\caption{Coherent perfect absorption wavefront injection in a one-port quarter bow-tie billiard at an arbitrary frequency of 8.8315 GHz. Output power ratio vs input power (lower axis, blue) and input phase (upper axis, red). The blue symbols correspond to the input signals power being swept at a fixed phase, and the red symbols correspond to the input signals phase being swept at a fixed power.}
\label{One_Port_CPA}
\end{figure}

The simplest demonstration of coherent perfect absorption is for a one-port scattering system, which only requires that a reflection zero scattering singularity exists at the real frequency of interest. In this case, any signal sent into the system is completely absorbed as long as the system remains linear. In Figure \ref{One_Port_CPA}, we first used the singularity on demand optimization method to create a $det(S)=0+i0$ condition at an arbitrary frequency of 8.8315 GHz. Then we demonstrate that for input signals with a range of over 70 dB of input power and $2\pi$ phase variation, we have generally over seven orders of magnitude of absorption. For the input power sweep the phase was fixed at 0 degrees, and for the input phase sweep the power was fixed at 0 dBm. As the input power decreases, the denominator of the output to input power ratio shrinks and the effective noise level increases, causing the power ratio to increase slightly. 

\begin{figure}[htb]
\centering
\includegraphics[width=8.9cm]{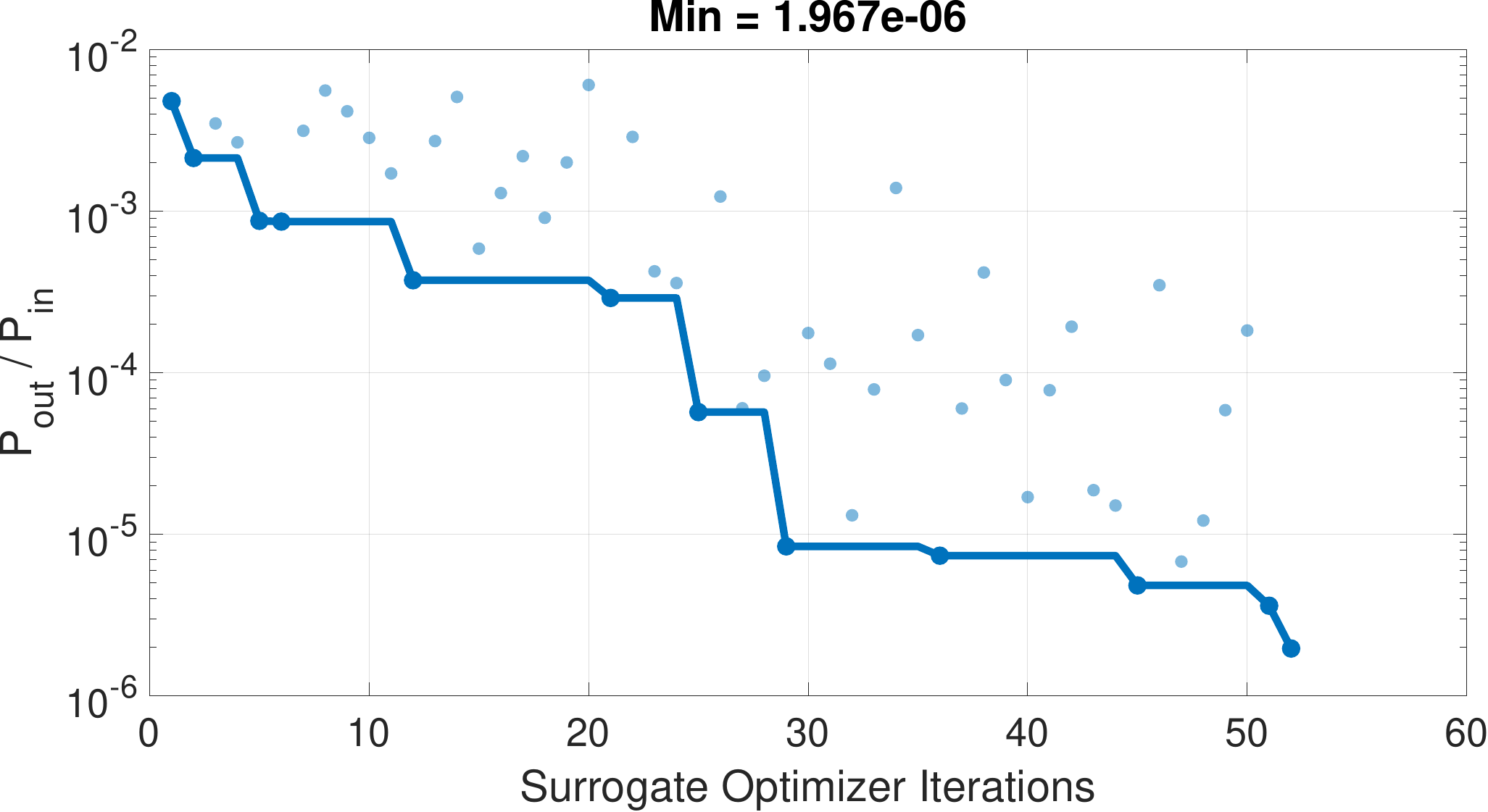}
\caption{Coherent perfect absorption wavefront injection optimization process in a three-port microwave graph at an arbitrary frequency of 9.6997 GHz. The points correspond to the value of the output to input power ratio at each optimization step.}
\label{1D_3_Port_CPA}
\end{figure}

\begin{figure}[htb]
\centering
\includegraphics[width=8.9cm]{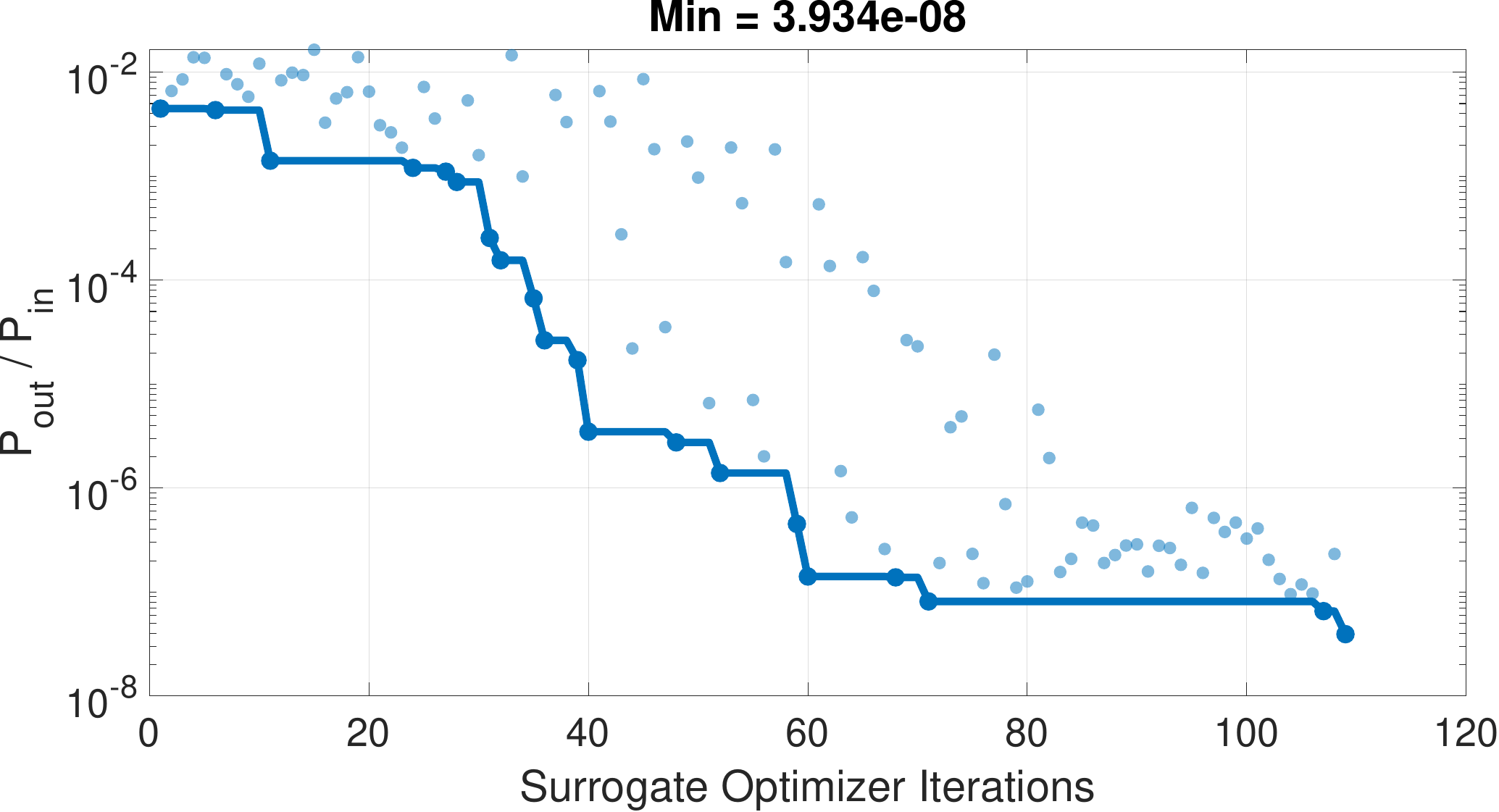}
\caption{Coherent perfect absorption wavefront injection optimization process in a two-port three dimensional cavity at an arbitrary frequency of 3.296091345 GHz. The points correspond to the value of the output to input power ratio at each optimization step.}
\label{3D_2_Port_CPA}
\end{figure}

In Figures \ref{1D_3_Port_CPA} and \ref{3D_2_Port_CPA}, we demonstrate coherent perfect absorption in a three-port microwave graph and a two-port three dimensional cavity, respectively, using an optimization process to find the CPA wavefront. In each of the figures, the output to input power ratio is plotted versus each iteration of the injection optimization process, and the solid line follows the cumulative minima throughout the iterations. These CPA injections were demonstrated at arbitrary frequencies using the singularity on demand optimization method to create a $det(S)=0+i0$ condition at 9.6997 GHz for the three-port graph, and 3.296091345 GHz for the two-port three dimensional cavity. By demonstrating the injection of the coherent perfect absorption wavefront in multiple scattering systems, we show that CPA can be achieved in generic scattering systems of any wave propagation dimension.

\begin{figure}[htb]
\centering
\includegraphics[width=8.9cm]{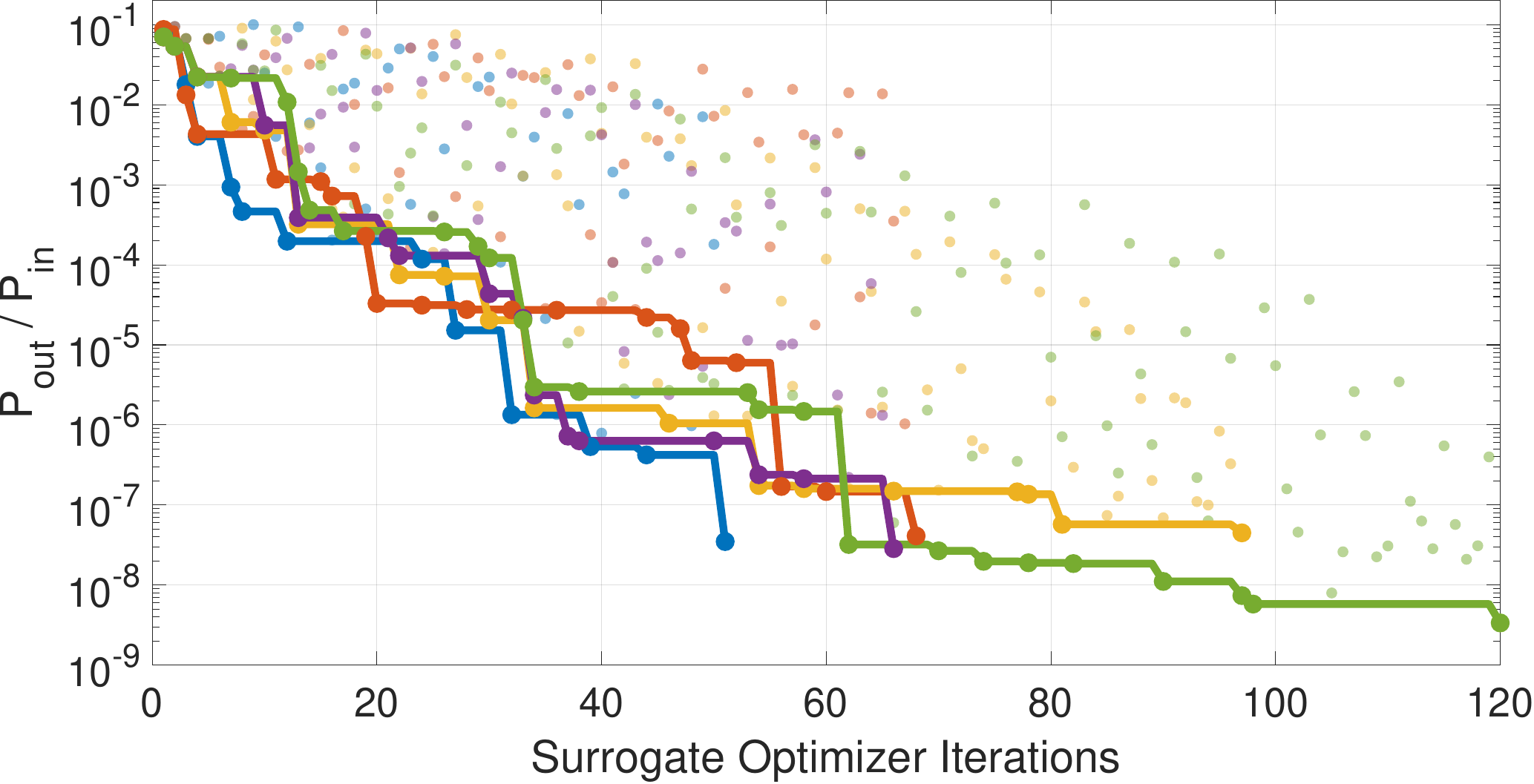}
\caption{Coherent perfect absorption wavefront injection optimization process in the two-port quarter bow-tie billiard using randomly initialized wavefronts at the start of each optimization. The points correspond to the value of the output to input power ratio at each optimization step, and the colors of the points correspond to optimizations with a different initial wavefront. The solid lines follow the cumulative minimum value through the optimization iterations. For these measurements we first created $det(S)=0+i0$ conditions at a frequency of 9 GHz using the singularity ``on demand'' optimization process.}
\label{Random_Start_2_Port}
\end{figure}

\begin{figure}[htb]
\centering
\includegraphics[width=8.9cm]{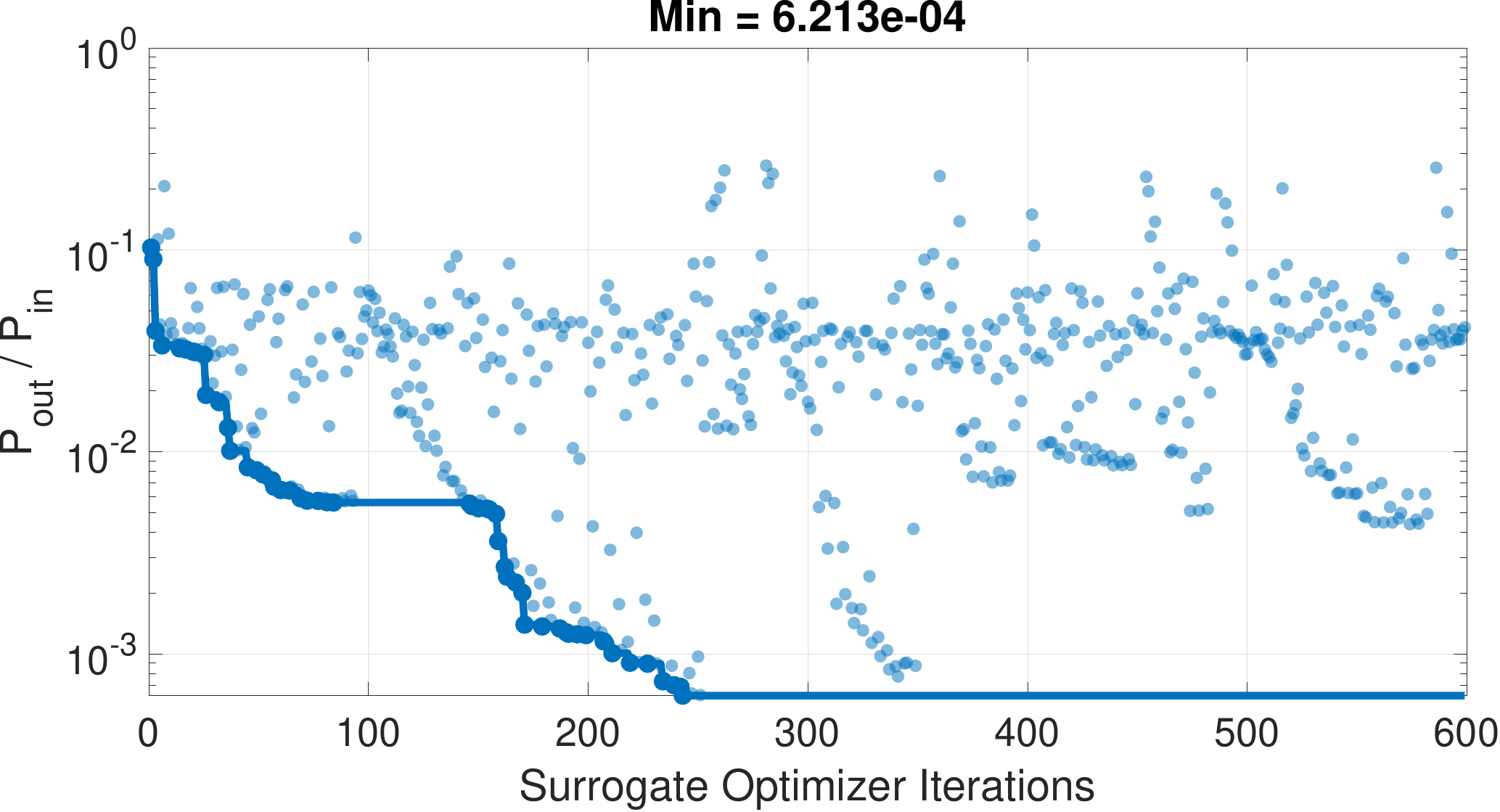}
\caption{Coherent perfect absorption wavefront injection optimization process in the four-port quarter bow-tie billiard using a randomly initialized wavefront. The points correspond to the value of the output to input power ratio at each optimization step. The solid line follows the cumulative minimum value through the optimization iterations. For this measurement we first created $det(S)=0+i0$ conditions at a frequency of 9 GHz using the singularity ``on demand'' optimization process.}
\label{Random_Start_4_Port}
\end{figure}

To demonstrate the ability of the optimization process to find the CPA wavefront from a randomly initialized wavefront instead of using the CPA eigenvector as an initial guess, we show Figures \ref{Random_Start_2_Port} and \ref{Random_Start_4_Port}. In both figures, we first created $det(S)=0+i0$ conditions at a frequency of 9 GHz using the singularity ``on demand'' optimization process. In Figure \ref{Random_Start_2_Port}, we show the optimization process for finding the two-port CPA wavefront for five different randomly initialized wavefronts. We see that the number of steps it takes to find a minimum value of the output to input power ratio can vary significantly, and the minimum value itself can vary a fair bit. Additionally, the starting value of $P_{out}/P_{in}$ for each of the optimizations is many orders of magnitude larger than shown in Figure \ref{Multi_Port_CPA}, which starts at a value of approximately $5 \times 10^{-7}$. In the two-port case, using the calculated CPA eigenvector as an initial guess is extremely useful to start in the vicinity of the ``true'' CPA wavefront. Knowing the CPA eigenvector greatly simplifies the optimization process as we can fix the power and phase of one of the microwave sources and adjust the other microwave source until the CPA wavefront is found. Additionally, from the CPA eigenvector information we can further increase the speed of finding the CPA wavefront by setting bounds on the amplitude and phase of the changing microwave source, reducing the parameter space the optimizer is allowed to explore. Due to the limited dynamic range of microwave sources, for the randomly initialized wavefront optimizations we didn't fix the power of one of the sources or set any bounds on the amplitudes and phases in case the other microwave source couldn't be adjusted enough to find the CPA wavefront. In Figure \ref{Random_Start_4_Port}, we show the optimization process attempting to find the four-port CPA wavefront from a randomly initialized input wavefront, but ultimately failing through 600 iterations. Using the knowledge of the CPA eigenvector is crucial to finding the CPA wavefront in higher port systems in a reasonable time frame. With that knowledge, the dimensionality of the optimization process can be reduced from $2M$ parameters ($M$ amplitudes and $M$ phases) to $2M-2$ parameters by fixing one of the microwave sources and the searchable parameter space of the other three microwave sources can be reduced by setting bounds on the amplitudes and phases.

\begin{figure}[htb]
\centering
\includegraphics[width=8.9cm]{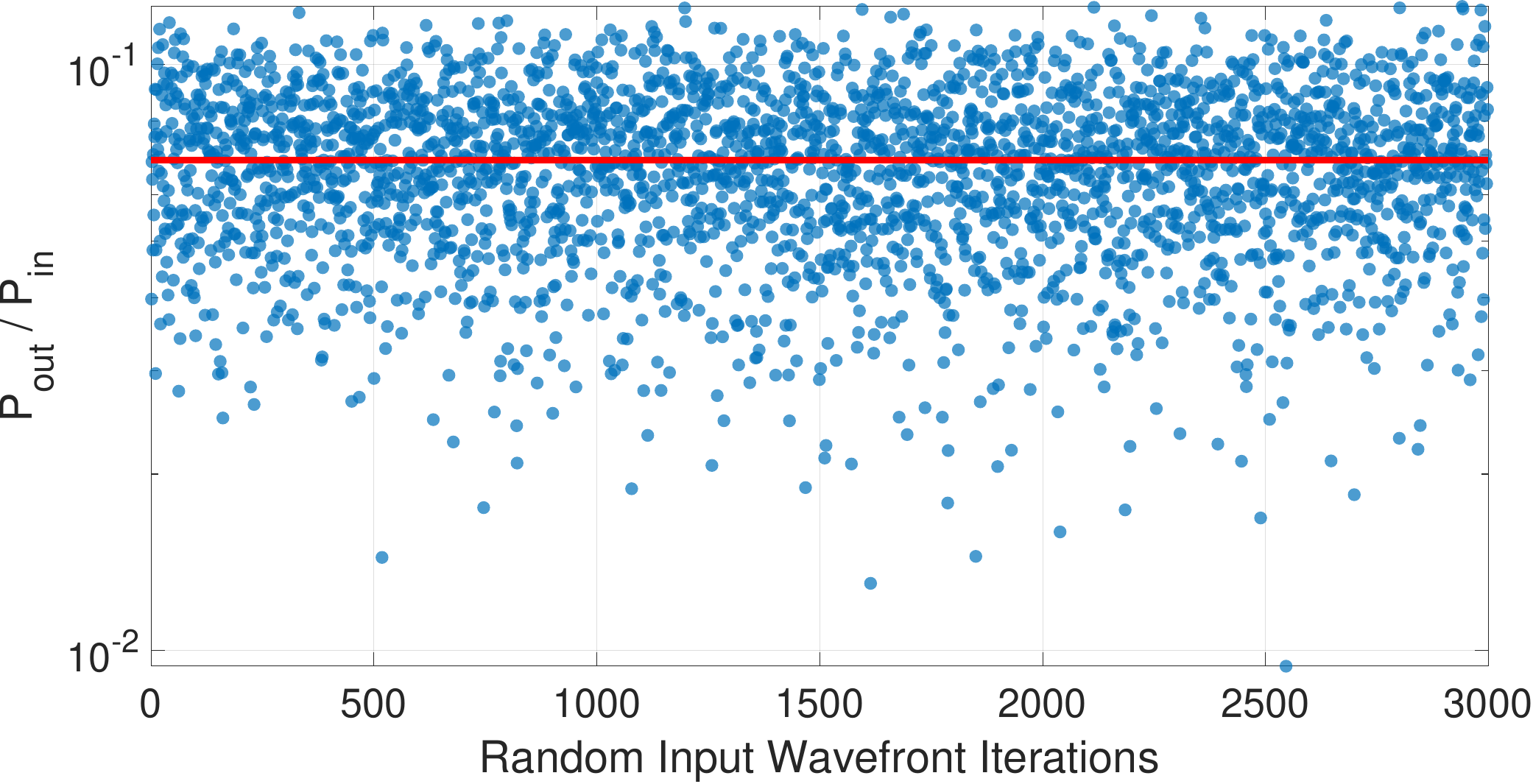}
\caption{Random wavefront injection into the four-port quarter bow-tie billiard. The points correspond to the value of the output to input power ratio for a sequence of randomly generated input wavefronts at a frequency of 9 GHz. The amplitudes of the wavefront were generated from a dirichlet distribution, where each component could vary over a power range of approximately 35 dB. The phases of each component were chosen from a uniform distribution from $-\pi$ to $\pi$. The red line indicates the mean value of the output to input power ratio over all iterations, $6.87\times 10^{-2}$.}
\label{Random_Injection_4_Port}
\end{figure}

In Figure \ref{Random_Injection_4_Port}, we show the output to input power ratio of 3000 randomly initialized wavefronts injected into the four-port quarter bow-tie billiard at a frequency of 9 GHz. The amplitudes of the microwave sources were generated from a dirichlet distribution, where the power could vary over a range of approximately 35 dB, and each phase was chosen from a uniform distribution between $-\pi$ to $\pi$. We see that the output to input power ratio of nearly all of the input wavefronts is between $10^{-1}$ and $10^{-2}$, where the average value over the ensemble is $6.87\times 10^{-2}$. This distribution of output to input power ratio values for random input wavefronts is representative of the quarter bow-tie system as a whole. The average $P_{out}/P_{in}$ values will change a bit for the quarter bow-tie with only two or three ports and for different frequency/cavity conditions (metasurface voltage values), but generally the output to input power ratio is around this order of magnitude for randomly injected wavefronts.

\section{Characterization of the Vector Network Analyzer Receivers and Microwave Sources}\label{sec.B}

In this appendix, we characterize the noise floor of the vector network analyzer (VNA) receivers used to measure the scattering matrix and inject arbitrary wavefronts into the experimental systems. We also characterize the stability of the microwave sources by measuring how a fixed input signal varies over time.

\begin{figure}[htb]
\centering
\includegraphics[width=8.9cm]{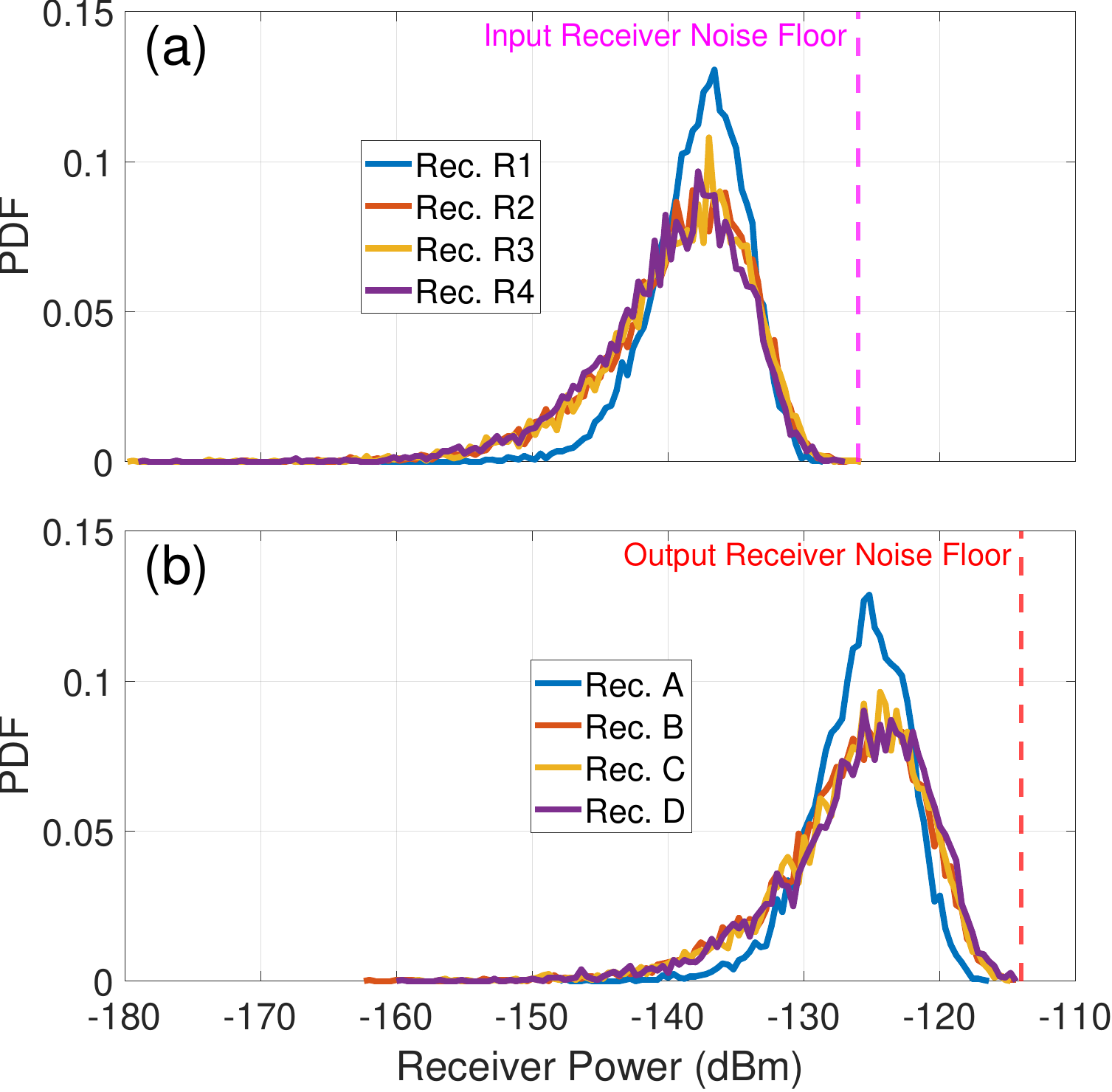}
\caption{Noise characterization of the vector network analyzer (VNA) receivers. The microwave sources were turned off for this measurement, while the receivers were set to measure at a frequency of 9 GHz. In this setup, the VNA was connected to the quarter bow-tie billiard and a dataset of approximately 6400 points was collected over three hours. (a) PDFs of the power measured by the input receivers (R1-R4) in dBm. The dashed red line at -126 dBm is the experimentally determined noise floor of the input receivers. (b) PDFs of the power measured by the output receivers (A-D) in dBm. The dashed red line at -114 dBm is the experimentally determined noise floor of the output receivers.}
\label{Receiver_Noise}
\end{figure}

In Figure \ref{Receiver_Noise}, the receivers of the VNA connected to the four-port quarter bow-tie billiard were set to measure ingoing and outgoing signals at a fixed frequency of 9 GHz over the course of three hours, while all four microwave sources were turned off. Approximately 6400 samples were collected, and the probability density functions (PDFs) of the powers measured by each receiver are plotted in dBm. The dashed lines at the larger side of the PDFs indicate the noise floor of the receivers, where we see that the input receivers have a noise floor of -126 dBm and the output receivers have a higher noise floor of -114 dBm. The receivers connected to ports 2-4 have approximately the same PDFs, while the receivers connected to port 1 are slightly different. This is due to port 1 containing extra internal VNA hardware options that are not present on the other ports.

\begin{figure}[htb]
\centering
\includegraphics[width=8.9cm]{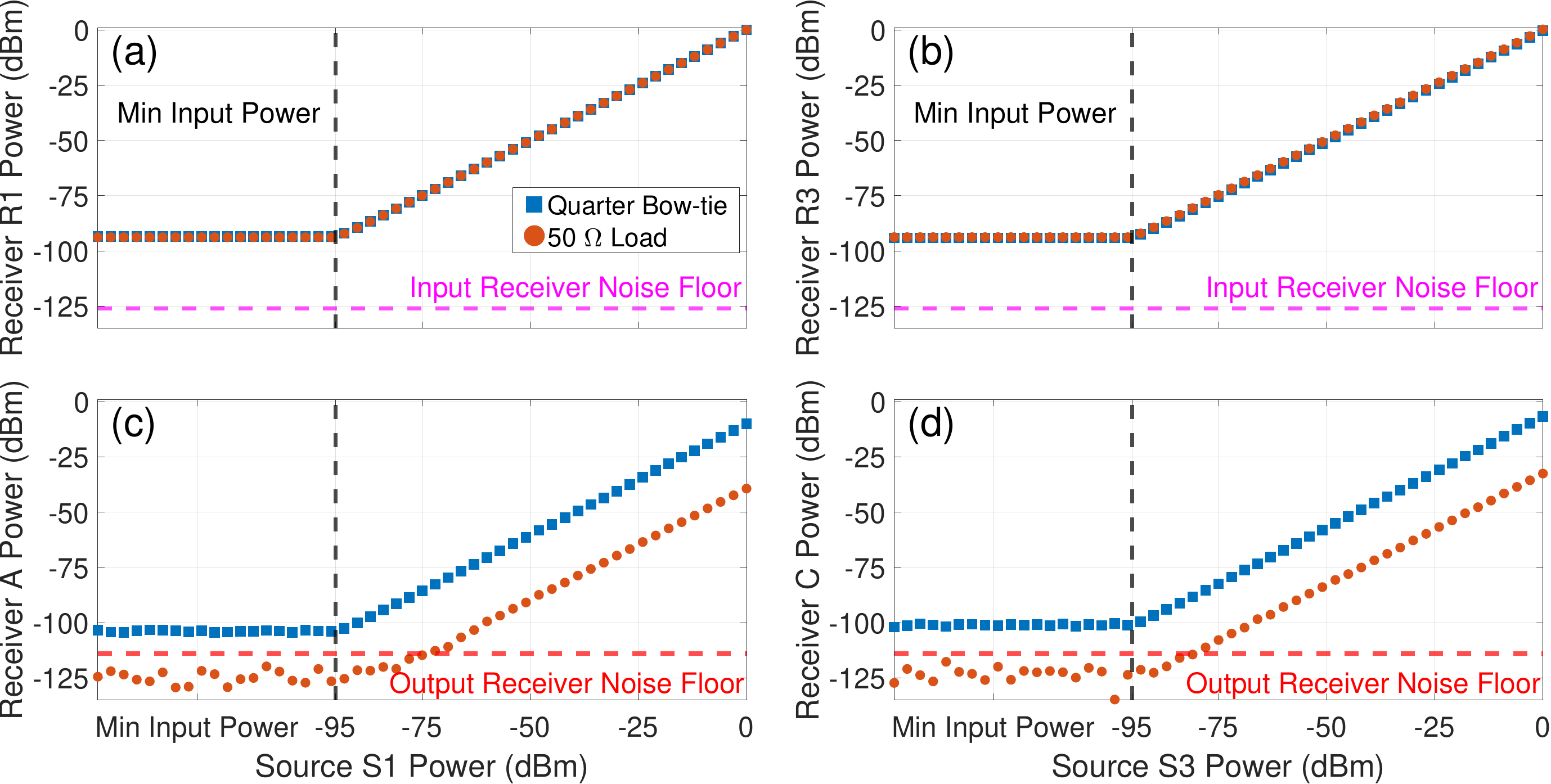}
\caption{Characterizing the measured power at the receivers of the vector network analyzer (VNA) as the source power is varied. The input receivers, R1 and R3, are shown in panels (a) and (b), and the output receivers, A and C, are shown in panels (c) and (d). The blue squares are the measured powers when the VNA was connected to the quarter bow-tie billiard at a frequency of 8.5 GHz, and the orange circles are the measured powers when the VNA was connected directly to 50$~\Omega$ loads at a frequency of 12 GHz. The minimum input power of sources S1 and S3 is -95 dBm, marked by the vertical dashed black lines. Left of the dashed black lines, the input power is held fixed at -95 dBm, as seen by the flatness of the measured powers of R1 and R3. The input (output) receiver noise floor determined from Fig.~\ref{Receiver_Noise} is marked with a pink (red) dashed horizontal line.}
\label{Input_Output_Char}
\end{figure}

In Figure \ref{Input_Output_Char}, we demonstrate that the measured power of the input and output receivers is linear in the source power down to the minimum input power of -95 dBm and that the measured powers above the noise floor are stable at the minimum input power. We measured the ingoing and outgoing powers when the VNA was connected to the two-port quarter bow-tie billiard at a frequency of 8.5 GHz (blue squares), and when the VNA was connected directly to 50$~\Omega$ loads at a frequency of 12 GHz (orange circles). The 50$~\Omega$ load measurement was done in order to show an example where the outgoing signals were below the noise floor of the receivers. Left of the dashed vertical black line indicating the minimum input power, the source power to the input receivers (R1 and R3) remained fixed at -95 dBm. At the minimal input power, the input receivers measure extremely consistent values as indicated by the flatness of the points left of the vertical dashed line in (a) and (b). For (c) and (d), when the output receivers were connected to the quarter bow-tie billiard (blue squares), the outgoing signal was above the noise floor and there is a little fluctuation in the measured signal, roughly characterizing the stability of the output receivers and system combination. When the output receivers were connected to the 50$~\Omega$ loads (orange circles), the outgoing signals hit the noise floor before the input power reaches its minimal value, and the measured power varies significantly after reaching the noise floor.

\begin{figure}[htb]
\centering
\includegraphics[width=8.9cm]{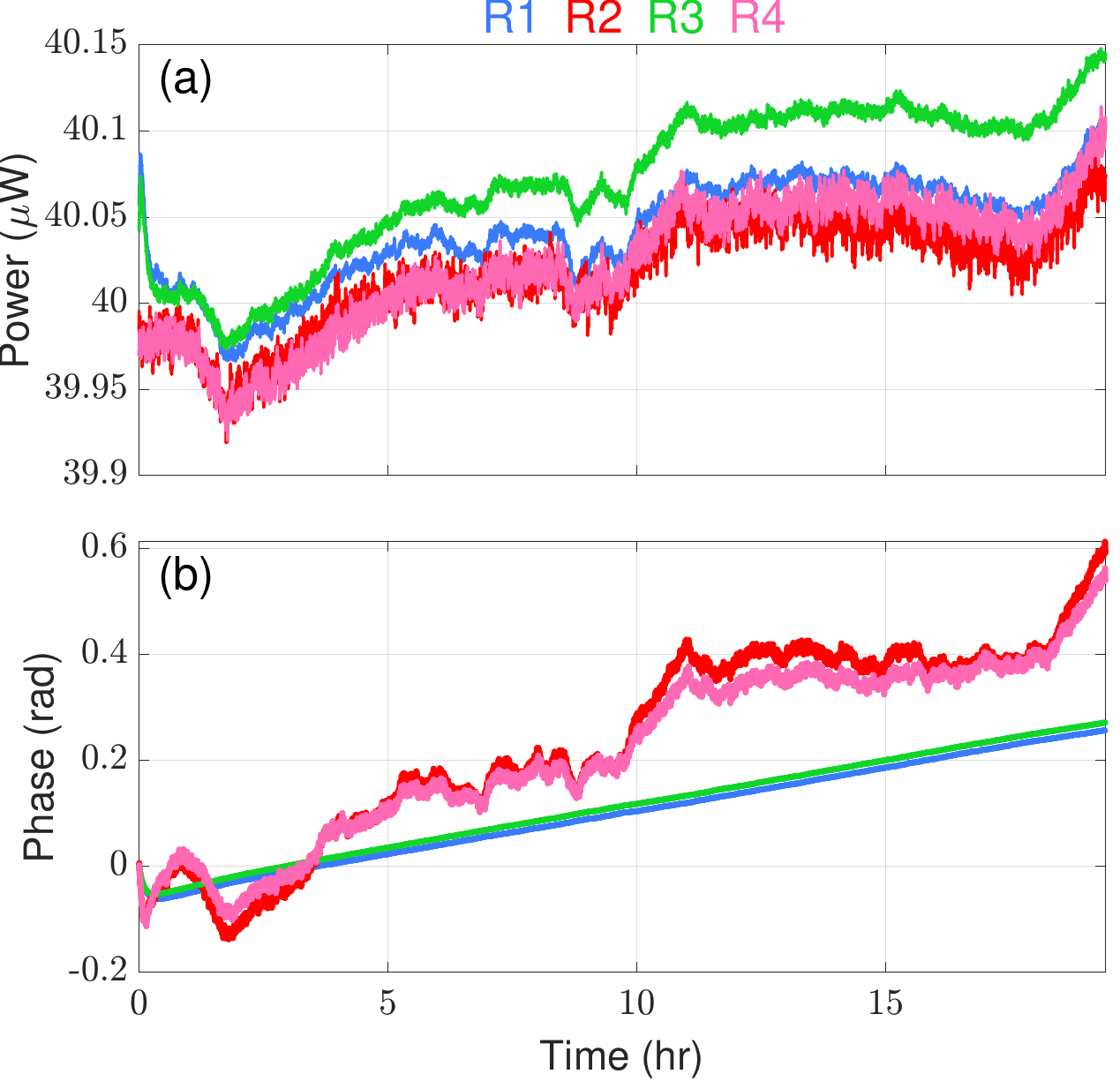}
\caption{Stability of microwave sources over time. For this measurement, all four microwave sources used in the four-port injection setup (see Fig. \ref{Receiver_Schematic}) were set to output a fixed power and phase at a frequency of 9 GHz, and the input receivers (R1-R4) measured how the input signals varied over time. The internal (external) vector network analyzer microwave sources were S1 and S3 (S2 and S4). (a) Input power of microwave sources measured by the receivers over time. Each curve corresponds to the same colored receiver indicated at the top of the plot. (b) Input phase of microwave sources measured by the receivers over time. Each phase curve was adjusted to start at zero for ease of comparison, and the colors of each curve correspond to the same receivers as indicated in (a).}
\label{Phase_Noise}
\end{figure}

In Figure \ref{Phase_Noise}, the stability of the microwave sources over time was measured. For this measurement, the microwave sources were set to inject a fixed signal at approximately the same power and phase at a frequency of 9 GHz. Before starting this measurement, the VNA and external microwave sources had been turned on for a few days, reducing instability associated with the equipment initially warming up. The input receivers of the VNA measured the ingoing signals over the course of approximately 19 hours. The internal microwave sources of the VNA (sources S1 and S3) correspond to the blue and green lines while the external microwave sources routed through the VNA (sources S2 and S4) correspond to the red and pink lines. The overall power drift in Figure \ref{Phase_Noise}(a) is similar for all sources, however the noisiness of the external source curves, as indicated by the thickness of the curves, is significantly larger than the internal sources. In Figure \ref{Phase_Noise}(b), while the phases of the internal microwave sources vary over time, they mainly vary coherently. This is in contrast to the external source phase curves which vary significantly more than the internal sources and incoherently. For demonstrating coherent perfect absorption, as the output to input power ratio shrinks, a fixed amount of noise in any system or input wavefront parameter will have an increasingly larger effect on the ability to reduce the output to input power ratio. This is caused by the sharpness of the output to input power ratio curve as a parameter approaches its optimal value corresponding to the CPA state. Figure 6 of Ref. \cite{Erb24} demonstrates the output to input power ratio sharpness in the two-port quarter bow-tie billiard with four plots of independent parameters being swept through their optimal CPA value while the non-swept parameters are held fixed at their optimal values.

\clearpage
\newpage

\bibliography{Bibliography.bib}

\end{document}